\documentclass[%
 reprint,
superscriptaddress,
showpacs,
 amsmath,amssymb,
 aps,
]{revtex4-1}

\usepackage{graphicx}
\usepackage{dcolumn}
\usepackage{bm}
\usepackage{siunitx}

\begin{document}


\title{Microscopic details of stripes and bubbles in the quantum Hall regime}

\author{Josef Oswald}
\email{Josef.Oswald@unileoben.ac.at}
\affiliation{%
Institut f\"{u}r Physik, Montanuniversit\"{a}t Leoben, Franz-Josef-Strasse 18, 8700 Leoben, Austria
}%

\author{Rudolf A.\ R\"{o}mer}
\email{R.Roemer@warwick.ac.uk}
\affiliation{
Department of Physics, University of Warwick, Coventry, CV4 7AL, UK%
}
\affiliation{
CY Advanced Studies  and LPTM (UMR8089 of CNRS),
CY Cergy-Paris Universit\'{e},
F-95302 Cergy-Pontoise, France%
}
\affiliation{
Department of Physics and Optoelectronics, Xiangtan University, Xiangtan 411105, Hunan, China
}

\date{\today}

\begin{abstract}
We use a fully self-consistent laterally resolved Hartree-Fock approximation for numerically addressing the electron configurations at higher Landau levels in the quantum Hall regime for near-macroscopic sample sizes. 
At low disorder we find, spatially-resolved, stripe- and bubble-like charge density modulations and show how these emerge depending on the filling factor. 
The microscopic details of these boundary regions determine the geometrical boundary conditions for aligning the charge density modulation either as stripes or bubbles. Transport is modelled using a non-equilibrium network model giving a pronounced anisotropy in direction of the injected current in the stripe regime close to half filling. We obtain a stripe period of $2.9$ cyclotron radii. Our results provide an intuitive understanding of its consequences in strong magnetic fields and indicate the dominance of many particle physics in the integer quantum Hall regime when studied at legnth scales .

\end{abstract}

\pacs{73.43.-f, 
      73.43.Nq, 
      73.23.-b 
}
\maketitle



In May 2019, the quantum Hall effect \cite{Klitzing1980a,Stormer1982} was formally included among the select group of high-precision experiments to form the basis of a new SI system of units based on the "Planck constant $h$, the elementary charge $e$, the Boltzmann constant $k$, and the Avogadro constant $N_A$" \cite{vonKlitzing2019}. This had long been awaited and certainly represents a great achievement and fitting fulfillment of the vision for "nat\"{u}rliche Masseinheiten" \cite{translation}
proposed by Max Planck \cite{Planck1900}. In his essay to celebrate this achievement \cite{vonKlitzing2019}, von Klitzing also points out "that a microscopic picture of the quantum Hall effect for real devices with electrical contacts and finite current flow is still missing." 
Prominent examples of such microscopic details are so-called "bubble" and "stripe" phases \cite{Fogler2002}. They have been identified, e.g., by transport experiments in higher Landau levels (LLs) of ultra-high mobility samples \cite{Lilly1999a,Du1999,Du2000b} and are characterized by strong transport anisotropies (stripes) or reentrance effects (bubbles). It is believed that the phases correspond to density modulations with characteristic geometric non-uniformities due to the interplay of Coulomb interaction and the wave functions in higher Landau levels.

Early work in modelling density modulations in the quantum Hall regime, starting from the celebrated Chklovskii, Shklovskii and Glazman picture \cite{Shklovskii1992}, assumed uni-directional charge-density waves (CDWs) \cite{Koulakov1995,Fogler1996d} while mean field treatments established the possibility of anisotropic phases in a Fermi liquid \cite{Fradkin1999,Spivak2006}. However, spatially resolved information does not yet exist of these phases. Experimentally this is due to the intrinsic challenge of using local scanning probes in low temperatures for such remotely doped systems \cite{Hashimoto2008,HasCFS12,Friess2014}. Nevertheless, much indirect experimental evidence for the existence of bubble and stripe phases has now been accumulated \cite{Kukushkin2011a,Liu2013a,Pan2014,Friess2014,Wang2015,Pollanen2015a,Msall2015b,Mueed2016a,Shi2017b,Friess2017b,Friess2018,Bennaceur2018b}. Theoretical modelling has likewise concentrated on transport signatures of these phases \cite{Ettouhami2006,Ettouhami2007,Cote2002,Cote2016} while spatially resolved models of bubbles and stripes are only available in clean systems \cite{Cote2003}. 
The period of the stripe patterns has been predicted to follow $d \propto R_c$ with $R_c= l_B \sqrt{2 n +1}$ the cyclotron radius in LL $n$ at the Fermi energy \cite{Goerbig2004a,Koulakov1995} and $l_B= \sqrt{\hbar/eB}$ the magnetic length.
Experimentally, $1.5 R_c$ \cite{Friess2014} and $3.6 R_c$ \cite{Kukushkin2011a} have been reported while $\sim 2.7 R_c$ is predicted theoretically \cite{Goerbig2004a}.
\begin{figure*}[t]
(a)\includegraphics[width=0.63\columnwidth,clip=true,trim=20 30 85 50]{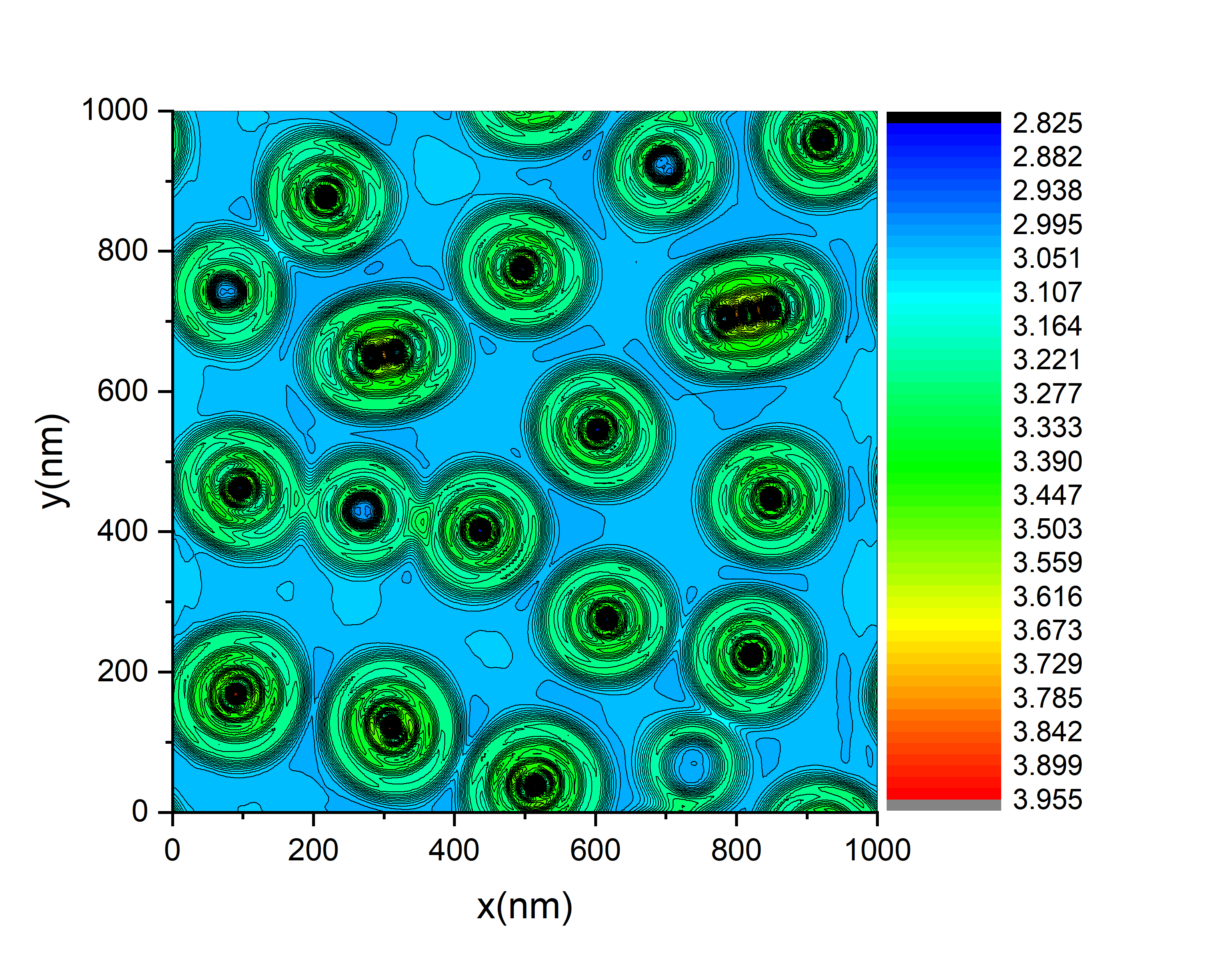} 
(b)\includegraphics[width=0.63\columnwidth,clip=true,trim=20 30 85 50]{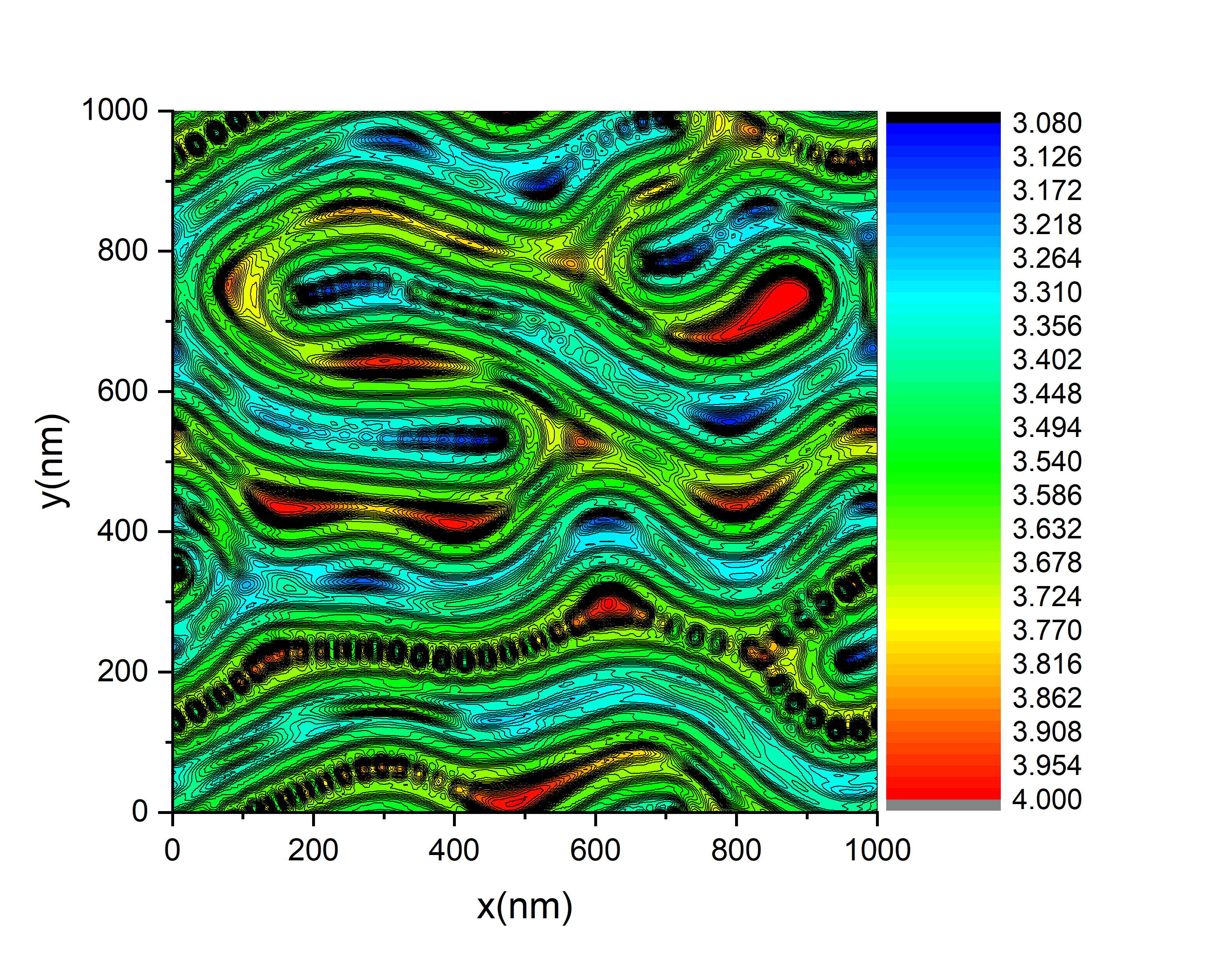} 
(c)\includegraphics[width=0.63\columnwidth,clip=true,trim=20 30 85 50]{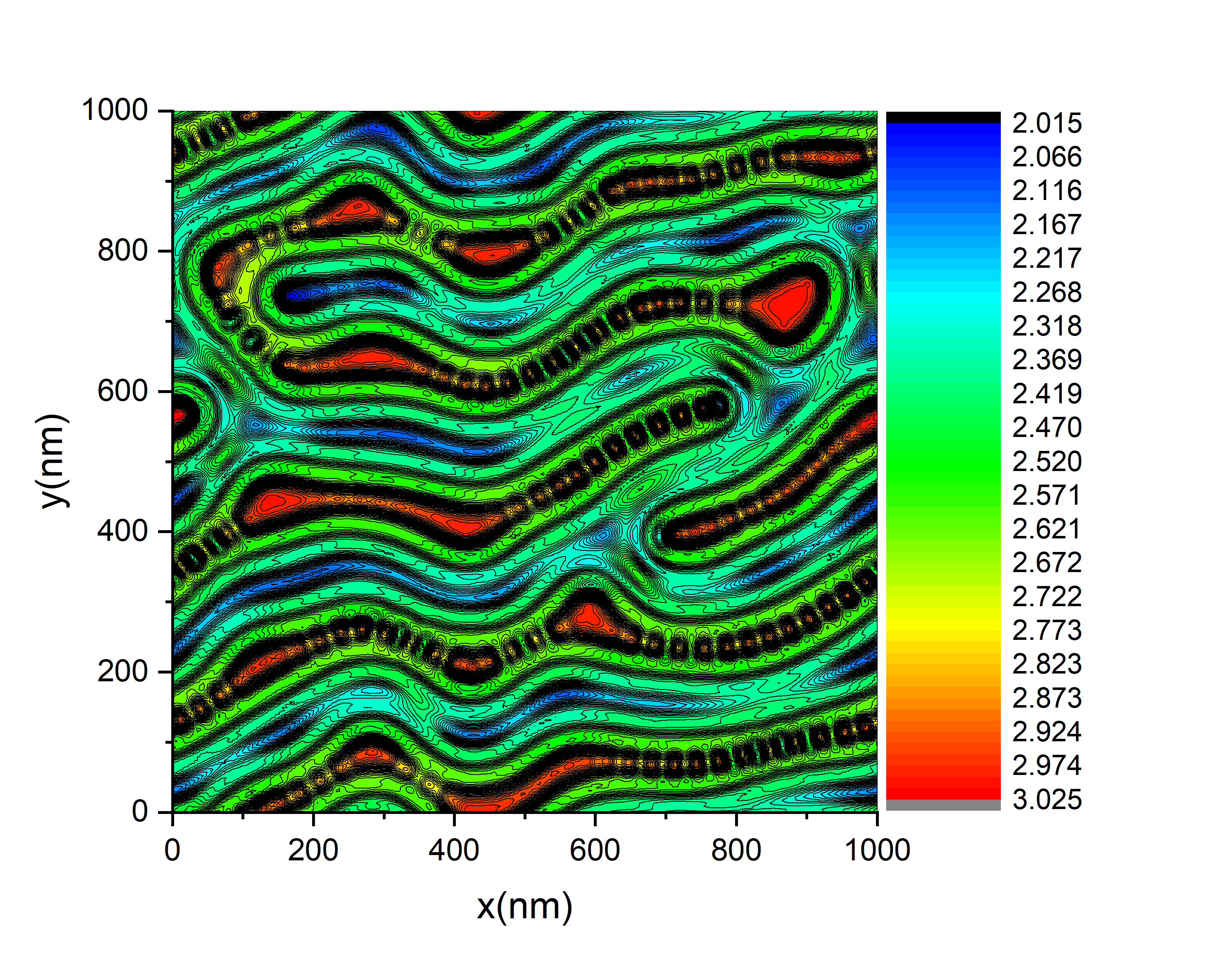} 
\caption{
\label{fig-nur}
\label{fig-bubble_big_nu_626}
\label{fig-stripes_big_nu65}
\label{fig-stripes_big_nu45}
Spatially resolved filling factor distribution $\nu_{\downarrow}(\vec{r})$  for different total filling factors (a) $\nu = 6.20$, (b) $\nu = 6.54$ and (c) $\nu = 4.54$ with $B=\SI{1.5}{Tesla}$. The colours denote different $\nu_{\downarrow}(\vec{r})$ values as indicated in the legends. The thin black lines are contours.}
\end{figure*}

In the present work, we show how stripes and bubbles emerge at weak disorder as self-consistent solutions of the Hartree-Fock (HF) equations, i.e.\ in the experimentally relevant regime and without any ad hoc assumptions beyond a smooth disorder. We provide the full spatial resolution of both phases from the length scale of $l_B$ to near macroscopic sample sizes. This high resolution allows \emph{quantitative} comparison with current experimental efforts \cite{Friess2014,Friess2017b,Friess2018,Mueed2016a,Kukushkin2011a}.
%
A central insight provided by our work is the importance of many-body aspects. It should be clear that a fully self-consistent HF approach in a disordered environment goes well beyond earlier Thomas-Fermi-based (non-)linear screening models. The inclusion of a converged exchange interaction term essentially alters the physics. This not only changes the spatial distribution of stripes and bubbles, but rather is the main reason of their emergence: neither pure Hartree nor a non-interacting model leads to emerging stripes/bubbles unless coupled with additional assumptions.
The key-mechanism is a Hund's rule behaviour for the occupation of the spin-split LLs. The resulting $g$-factor enhancement is then a local quantity depending on the local filling factor $\nu(\vec{r}) = 2 \pi l_B^2 \rho(\vec{r})$ \cite{OswaldEPL2017,OswaldPRB2017} with $\rho(\vec{r})$ the local carrier density. This \emph{exchange-enhanced} $g$-factor is a concept that allows to discuss the exchange interaction within the single electron picture \cite{Wiegers1997,Kendirlik2013}. By doing so, we find that the local variation of the enhancement of the Zeeman energy due to $\nu(\vec{r})$  has to be considered in addition to the laterally varying Hartree potential, leading to a modified effective potential for the electrons that also strongly modifies the screening behaviour. In addition to the largely repulsive Hartree part of the self-consistent Thomas-Fermi screening, the $\nu(\vec{r})$ dependence of enhanced Zeeman energy leads to a positive feedback loop in the self-consistent carrier redistribution and produces an instability of the electron density $\rho$, which may lead to jumps either to a locally full or locally empty LL \cite{OswaldEPL2017,OswaldPRB2017}, resulting in a clustering of the filling factor that is triggered by the disorder or edge potential. The boundaries of those clusters finally create narrow channels that align mainly along the edge or random potential fluctuations. In case of a very clean high mobility electron system such a trigger effect for the cluster formation by the random potential is missing and the electron system has to find such a cluster structure by self organization, resulting in the formation of stripes or bubbles.


The numerical simulations are performed as described in Refs.\ \cite{OswaldEPL2017,OswaldPRB2017,Sohrmann2007} via a variational minimization of the self-consistent Hartree-Fock-Roothaan equation \cite{Aok79,YosF79,MacA86,MacG88}. To simulate ultra-high mobility samples, the random potential strength is kept  low. The filling factor and the lateral system size are made as large as possible with respect to the available computing power. Configurations up to $\SI{1}{\mu m^2}$ are achievable as shown in Fig.\ \ref{fig-bubble_big_nu_626} with spatial resolution of $\sim \SI{4.4}{nm}$ well below $l_B$ for a magnetic field $B$ varying from, e.g., $1$ ($l_B\sim \SI{26}{nm}$) to $\SI{6.5}{Tesla}$ ($l_B\sim \SI{10}{nm}$). The random potential is generated by Gaussian impurity potentials of radius $\SI{40}{nm}$, the number of impurities is $N=2000$, their random placement results in a fluctuating potential of $V_\text{max} = \SI{0.43}{mV}$ and $V_\text{min} = \SI{-0.50}{mV}$ \cite{supplement}. At a total filling factor of, e.g., $\nu = 6.54$ ($= \nu_\downarrow + \nu_\uparrow = 3.54 + 3.0$) and $B = \SI{1.5}{Tesla}$, this corresponds to more than $2000$ electrons. In order to generate transport data, we employ a non-equilibrium network model (NNM) introduced previously \cite{OswO06}. A very large number of step by step calculations are required \cite{SohOR09} in the NNM. Hence, for keeping within the available computing time, the transport simulations have been performed for smaller sample size such as $500 \times \SI{500}{nm^2}$. 



In Fig.\ \ref{fig-bubble_big_nu_626} we depict the variation in $\nu_{\downarrow}(\vec{r})$ for three different densities at fixed magnetic field.
Fig.\ \ref{fig-bubble_big_nu_626}(a) shows the situation far from any half-odd average $\nu_{\downarrow}$.
The required area of the filled (spin-down) clusters is only a minor part of the total area. As seen in the figure this can be achieved by a nearly evenly-spaced distribution of "bubble"-like shaped clusters. Since we are considering already the $4$th partly-filled spin-split LL, the boundaries of the bubbles consist of three sub-stripes because of the $3$ nodes of the Landau basis function for the $4$th LL \cite{HasCFS12}. A boundary with such an internal structure takes up a substantial area as well and even tends to dominate the region of a single bubble as a whole. In other words, the area that one bubble needs is dominated by the width of the boundaries and not by the region of an assumed idealized full LL. 
When increasing the filling factor towards half filling, as shown in Fig.\ \ref{fig-stripes_big_nu65}(b), it becomes clear that such a bubble-like geometry of the clusters is impossible to achieve with the given total $\nu$, because round bubbles would leave too much unfilled space between the bubbles even if touching each other. The only way to remove the unfilled space between bubbles is a change of the geometry so that the boundaries of different clusters arrange almost in parallel, which means a transfer to a "stripe"-like geometry. 
In order to demonstrate that the width of the boundaries of the stripes and bubbles depend on the LL index, Fig.\ \ref{fig-stripes_big_nu65}(c) shows the half-filled LL at filling factor $\nu \approx 4.5$ instead of $\nu = 6.5$ and one can see that there is one sub-stripe less at the boundary, allowing a more dense arrangement of stripes or bubbles than in the higher LLs. 
These results suggest that the higher the filling $\nu$ the higher the tendency for creating the stripe pattern near a half-filled LL. Overall, the figures show that our HF calculations can provide a fascinating insight in the spatial behaviour of bubble and stripe configurations.

\begin{figure}[tb]
\includegraphics[width=\columnwidth,clip=false,trim=40 20 100 50]{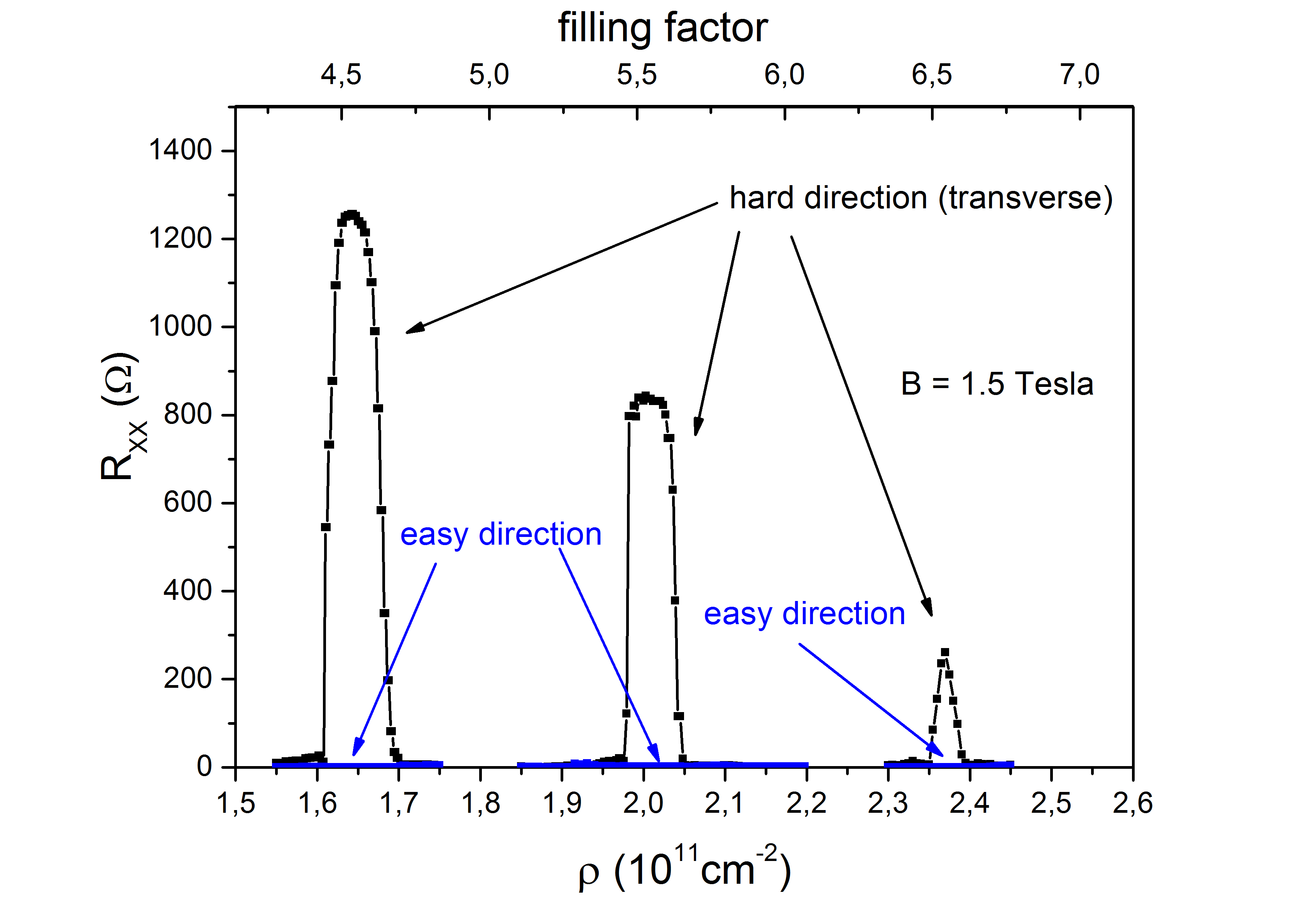}
\caption{
\label{transport}
Longitudinal resistance $R_{xx}$ for easy (horizontal) and hard (vertical) direction (cp.\ Fig.\ \ref{fig-nur}) as a function of carrier density $\rho$. 
The sample size is $500\times\SI{500}{nm^2}$, $B = \SI{1.5}{Tesla}$, the impurity radius $\SI{40}{nm}$, and the number of impurities $N=100$. The random impurity placement results in a fluctuating potential from $V_\text{min} = \SI{-1.7}{mV}$ to $V_\text{max} = \SI{1.7}{mV}$.}
\end{figure}

Fig.\ \ref{transport} shows the transport data obtained by applying the NNM at different carrier densities. 
Since the stripe alignment tends to be more horizontal, i.e.\ along the $x$-direction as shown in Fig.\ \ref{fig-nur}, the longitudinal resistance $R_{xx}$ appears higher for vertical sample current and lower for horizontal current flow \cite{supplement}. This is consistent with experimental observations resulting in large $R_{xx}$ peaks for vertical current, while for a horizontal current the $R_{xx}$ peaks are hardly visible. When the difference between horizontal and vertical $R_{xx}$ is no longer prominent, we find that we have reached a bubble phase.
%

\begin{figure}[thb]
(a)\hspace*{0ex}
\includegraphics[angle=0,keepaspectratio=true,width=0.95\columnwidth,clip=false,trim=20 140 650 0]{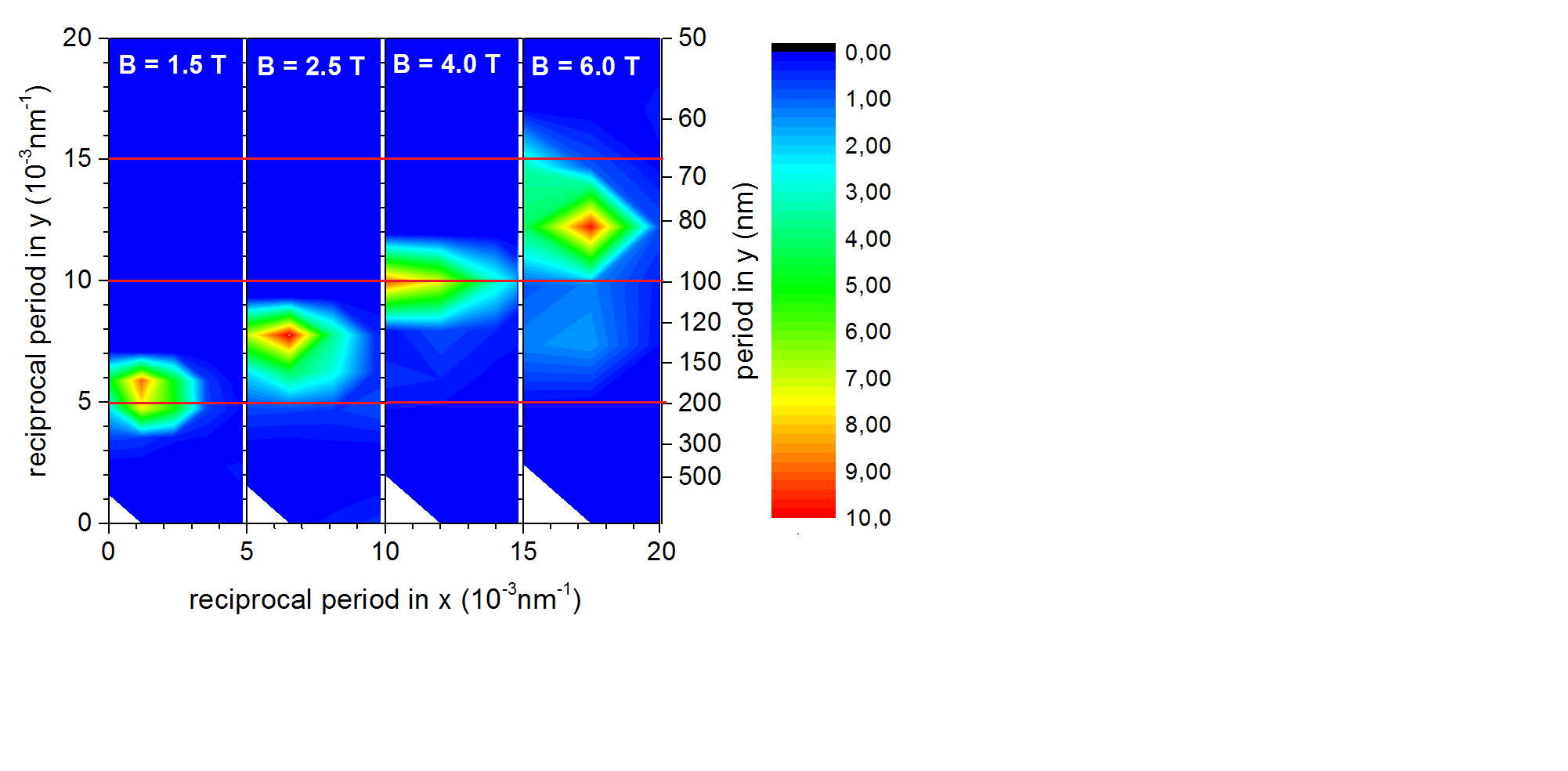}\\
(b)\includegraphics[width=0.95\columnwidth,clip=true]{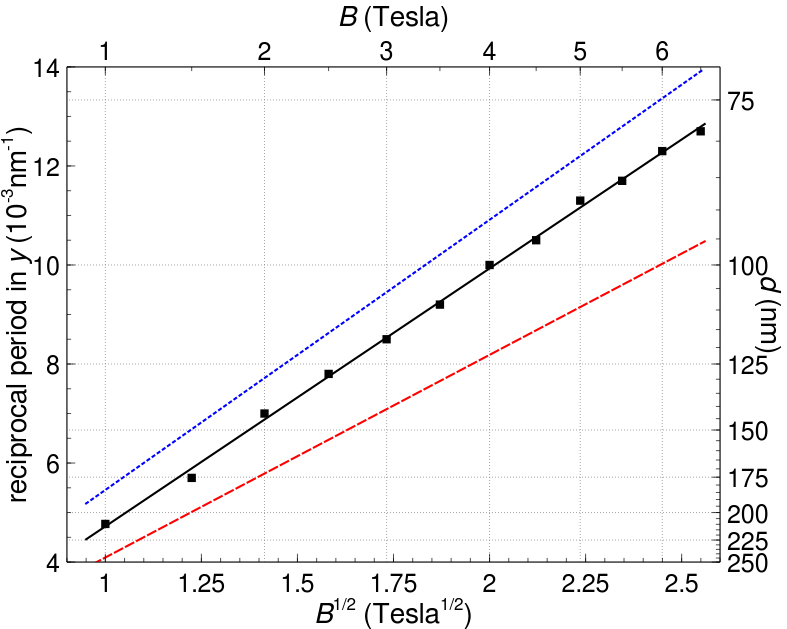}
\caption{\label{fig-fourier}\label{fig-FFT}
(a) Two-dimensional Fourier spectra at $\nu = 4.5$ for different $B$ fields. The different spectra are shifted to the right for better visibility. 
For a sample size of $1000 \times \SI{1000}{nm^2}$, the frequency in units of $\SI{0.001}{nm^{-1}}$ (left axis) equals the number of stripes that can be accommodated within the boundaries. The corresponding period length is shown on the right axis and the color bar indicates the Fourier intensity in arb.\ units.
The horizontal red lines incidate selected values.
(b) Variation of the reciprocal stripe period $1/d$ ($\blacksquare$) as function of $\sqrt{B}$. The solid lines corresponds to $d = \alpha \sqrt{2 n + 1}R_c$ for LL index $n=3$ with $\alpha=2.7$ (blue line, theory) \cite{Goerbig2004a}, $3.6$ (red line, experiment) \cite{Kukushkin2011a} and $2.9$ (black line, our fit), respectively. The grey dotted grid lines highlight selected $B$ and $d$ values.}
\end{figure}

The characteristics of the microscopic structure of the stripes consist on one hand in the periodicity of the stripe pattern and on the other hand in the microscopic details of the boundaries of the stripes. Together with the geometric shape of the stripes (and, indeed, the bubles), these determine the relation between the areas of full and empty LLs. 
Focusing first on the periodicity, Fig.\ \ref{fig-fourier} presents the 2D-Fourier transformation of the lateral carrier density at various magnetic fields from $B=\SI{1}{Tesla}$ to $B=\SI{6.5}{Tesla}$ at a fixed filling factor of $\nu = 4.5$. 
While Fig.\ \ref{fig-fourier} (a) shows only a selection of a few spectra, Fig.\ \ref{fig-fourier} (b) displays the trend of all evaluated spectra. As can be seen in $x$-direction there is a smooth and somewhat smeared out  distribution that starts from zero, indicating that there is no clear periodicity in $x$-direction. In contrast, in $y$-direction, there is a well pronounced maximum which matches the reciprocal stripe period in $y$-direction as also seen in Figs.\ \ref{fig-stripes_big_nu65} (b+c). For $B=\SI{1.5}{Tesla}$ the reciprocal period (wave number) appears to be close to $6$ \cite{scale} and for $B=\SI{6}{Tesla}$ it matches $12$, which is consistent with a $\sqrt{B}$ dependence. Furthermore, we extract the mean period of the corresponding stripe patterns, similar to those shown in Fig.\ \ref{fig-nur}, to be approximately $\SI{175}{nm}$ for $B=\SI{1.5}{Tesla}$ and $\SI{81}{nm}$ for $B=\SI{6}{Tesla}$. Extending the analysis to other $B$ values, we indeed find a clear $\sqrt{B}$ behaviour as shown in Fig.\ \ref{fig-fourier} (b).
Previous theoretical results on the period $d$ of the stripes have led to the expectation $d = \alpha R_c =\alpha \sqrt{(2n+1)\hbar B/e}$  
\cite{Goerbig2004a,Koulakov1995}. Experimentally, $\alpha=1.5$ \cite{Friess2014} and $3.6$ \cite{Kukushkin2011a} have been estimated. For $n=3$ and $B=\SI{1.5}{Tesla}$, these give $\SI{88}{nm}$ and $\SI{210}{nm}$, respectively. Obviously, only the latter one is compatible with our result at $\SI{1.5}{Tesla}$. Conversely, from Fig.\ \ref{fig-fourier} (b), we extract a value $\alpha=2.9 \pm 0.1$. This agrees very well with previous straight-line CDW-based predictions of $2.7$ \cite{Goerbig2004a} and $2.8$ \cite{Fogler1996d}.



\begin{figure}[bt]
(a)\hspace*{0ex}\includegraphics[width=0.34\columnwidth,clip,trim=120 40 60 60]{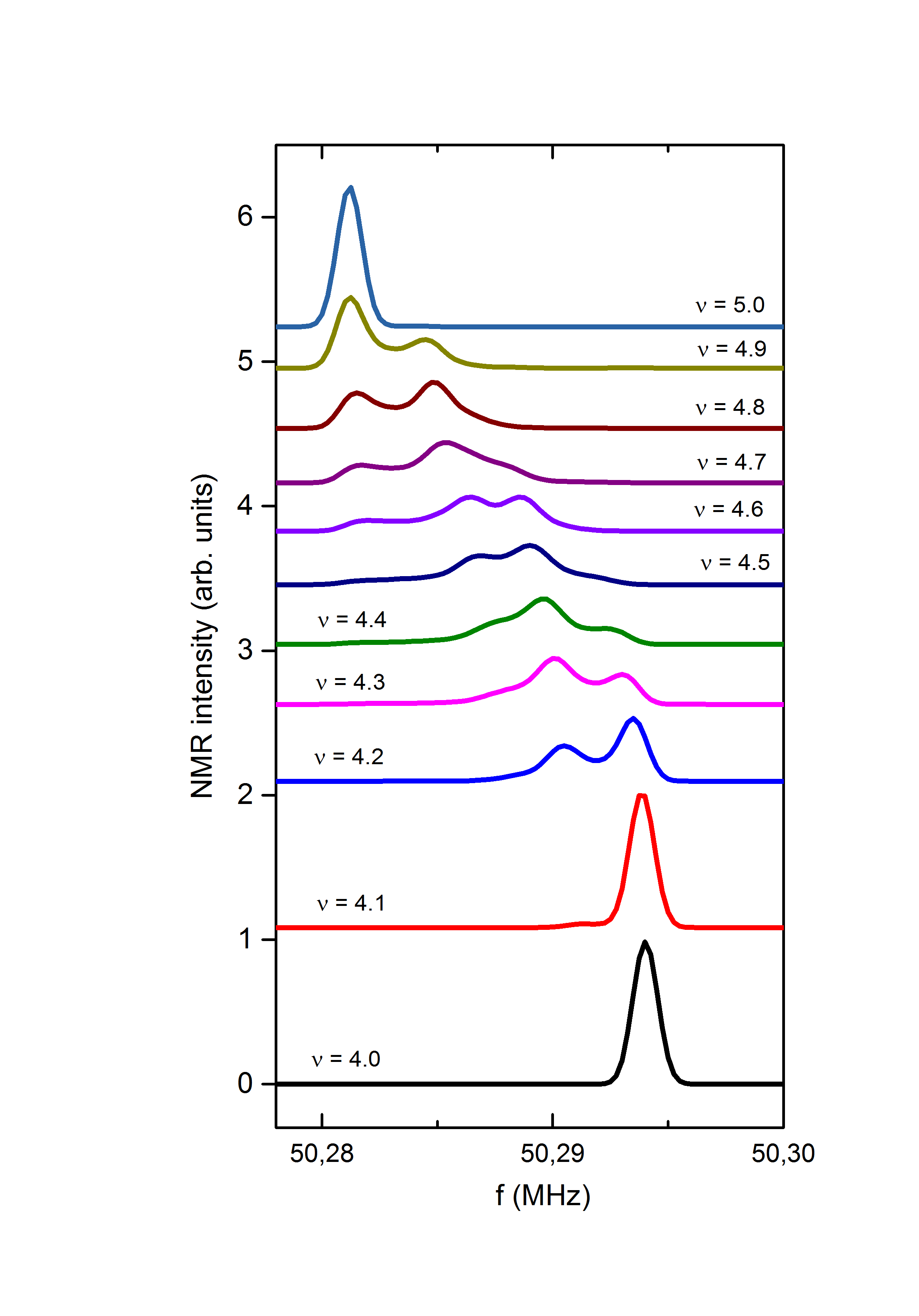}
(b)\vspace*{-1ex}\hspace*{0ex}\includegraphics[width=0.54\columnwidth,clip=true,trim=20 70 0 0]{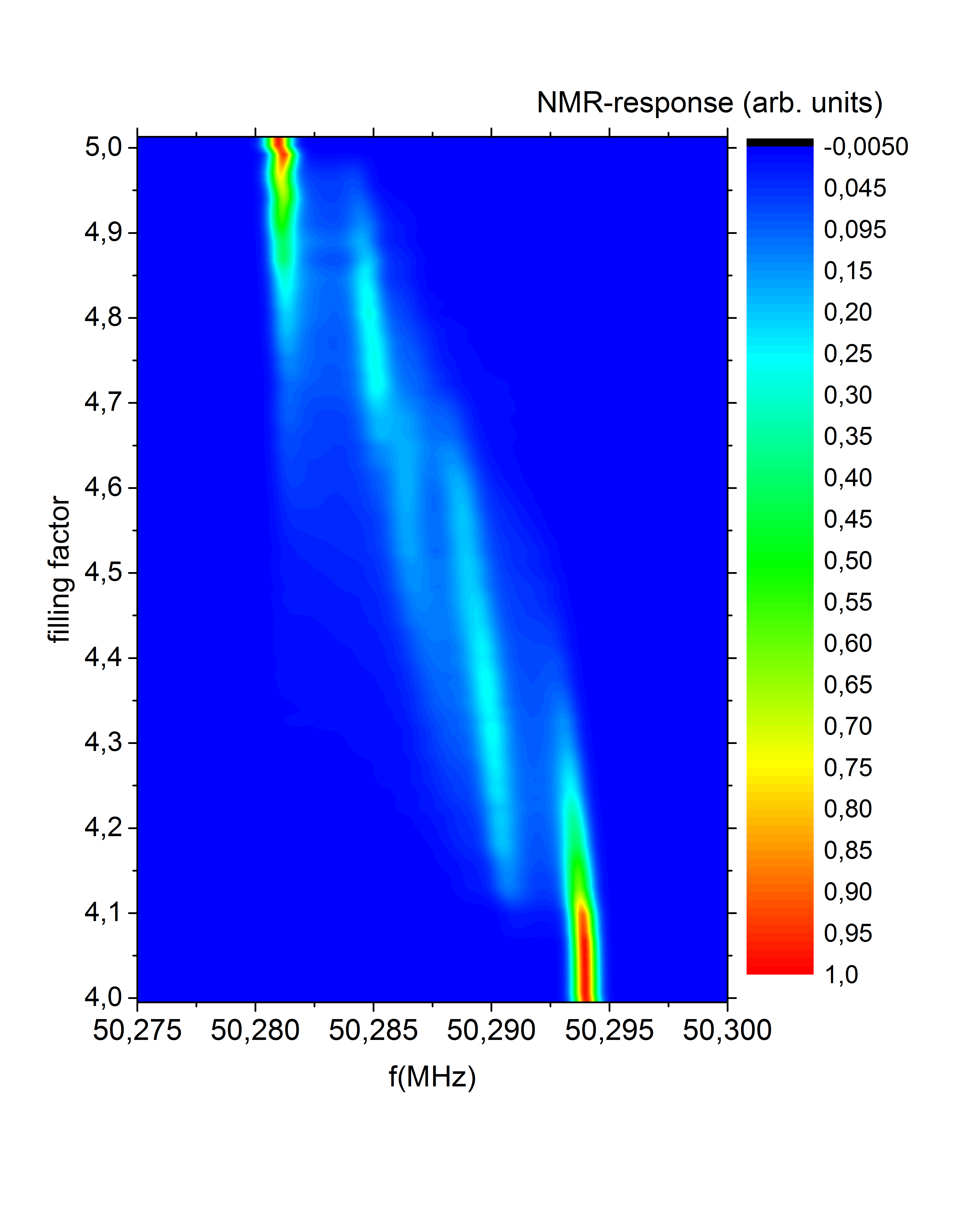} \\
\caption{
\label{fig-NMR}
(a) Simulation results of $I_{\nu}(f)$ using our HF results for $\nu_{\downarrow}(\vec{r})$ as input. The traces are shifted vertically for better visibility and varied in steps of $\Delta\nu=0.1$. 
(b) Color plot of $I_{\nu}(f)$ for very fine resolution $\Delta\nu=0.01$. The color legend gives the numerical values of $I_{\nu}(f)$.
}
\end{figure}

Friess et al.\ \cite{Friess2014} investigated the Knight shift in nuclear magnetic resonance (NMR) spectra to demonstrate the co-existence of regions with different spin polarization due to the periodic variation of the filling factor in stripe and bubble phases. Their method provides direct information about the area fractions, although there is no direct information about geometry and periodicity. In order to extract also microscopic details of the structural information, they use a semiclassical model \cite{Tiemann2012,Friess2014} 
based on superpositions of the \emph{single} electron densities obtained from the Landau basis functions. 
%
In our case, we can similarly model the NMR intensity as
$I_{\nu}(f)=\int \mathcal{G}\{f-[f_{0,\nu-1}-\nu(\vec{r})\cdot K_{\text{max},\nu}]\})\ dr^2$ with Gaussian $\mathcal{G}$ describing the absorption spectrum of individual nuclei \cite{Tiemann2012}, $f$ the NMR frequency, 
$K_\text{max}$ the maximal Knight shift for the fully spin-polarized LL at odd $\nu$ and $f_0$ the frequency of the non-shifted NMR line for the non spin-polarized situation at even filling factor $\nu-1$. Numerically, $I_{\nu}(f)$ is calculated by evaluating our \emph{interacting} $\nu(\vec{r})$ at each $\vec{r}$ and summing over all points of a typically $229 \times 229$ grid. 
%
In Fig.\ \ref{fig-NMR}, we show that $I_{\nu}(f)$ exhibits features that red-shift to lower frequencies when increasing $\nu$, e.g., from the non-spin polarized $\nu=4$ to a fully spin polarized $\nu=5$ (see supplement \cite{supplement} for $\nu=2 \rightarrow 3$). 
%
At intermediate filling factors the spectral response splits, mainly into double/triple peak structures. This indicates the co-existence of regions with different filling factor as expected from a stripe or bubble like electron distribution (see supplement \cite{supplement}). Around $\nu \sim 4.4$ one can recognise  \emph{three} peaks with changing weights in the superposition while varying the total filling factor. This is in good agreement with the experimental results \cite{Friess2014}, but naturally less so when compared to a single-particle modelling.

Intuitively, the driving force for the formation of bubbles and stripes can be understood as follows:
the Hund's rule behaviour causes a $g$-factor enhancement that maintains the tendency to fill up just one spin-level as much as possible before starting to fill up the next spin-level. When entering a stripe/bubble from outside (e.g.\ $\nu = 4.0$, not spin polarized) to inside (e.g.\ $\nu = 5.0$, fully spin polarized) the $g$-factor enhancement also drastically increases. Inside the stripe/bubble the spin-levels are strongly pushed apart, corresponding to a significantly lower energy of the occupied lower spin state, while outside there is minimal $g$-factor enhancement at even filling factor $\nu = 4.0$. In terms of an effective single particle picture the electrons in the stripes/bubbles encounter a potential well established by the local variation of the $g$-factor enhancement and which dominates over the repulsive Hartree interaction. From a qualitative point of view, the stripes can be understood as leaky non-coupled electron wave guides that consist of self assembled one-dimensional potential wells and the leakage shows up as the boundary region as discussed before. 
%


In conclusion, we have computed the microscopic picture of stripe and bubble phase for the IQH regime at high LLs and in weak disorder. Our results rely on spatially-resolved, self-consistent HF calculations of nearly macroscopic sizes of $\mathcal{O}(\SI{1}{\mu m^2})$ and, for transport calculations, are coupled to device contacts with finite currents. 
The existence of microscopic stripes and bubbles at weak disorder is thus confirmed. Their spatial features show intriguing extended oscillations along the stripes and surrounding the bubbles. These are clearly due to the structure of the underlying Landau states.
We find that the stripe period scales with ${B}^{-1/2}$ as expected and agrees in detail with previous experimental measurements. 
Overall, together with results from HF calculations in the strong disorder regime \cite{Sohrmann2007,OswaldPRB2017,OswaldEPL2017}, this shows that the IQH regime can now be described spatially resolved with high accuracy from microscopic to near-macroscopic length scales. Our results shine new light on the understanding of the microscopic picture of the IQH effect 
and demonstrate the permanent dominance of many particle physics for quantum Hall physics. The demonstrated Hund's rule behavior in context with the $g$-factor enhancement allows to incorporate the exchange interaction into an intuitive understanding of the major effects driving weakly and strongly disordered quantum Hall systems.





This work received funding by the CY Initiative of Excellence (grant "Investissements d'Avenir" ANR-16-IDEX-0008) and developed during RAR's stay at the CY Advanced Studies, whose support is gratefully acknowledged. We thank Warwick's Scientific Computing Research Technology Platform for computing time and support. UK research data statement: Data accompanying this publication are available from the corresponding authors.


\bibliographystyle{apsrev4-1}

\clearpage
\part*{Supplemental Material}
\def\thefigure{S\arabic{figure}}
\setcounter{figure}{0}
\def\thetable{S\arabic{table}}
\def\theequation{S\arabic{equation}}


\title{\textit{Supplemental Material for}\\%
Microscopic details of stripes and bubbles in the quantum Hall regime}

\author{Josef Oswald}
\email{Josef.Oswald@unileoben.ac.at}
\affiliation{%
Institut f\"{u}r Physik, Montanuniversit\"{a}t Leoben, Franz-Josef-Strasse 18, 8700 Leoben, Austria
}%

\author{Rudolf A.\ R\"{o}mer}
\email{R.Roemer@warwick.ac.uk}
\affiliation{
Department of Physics, University of Warwick, Coventry, CV4 7AL, UK%
}
\affiliation{
CY Advanced Studies  and LPTM (UMR8089 of CNRS),
CY Cergy-Paris Universit\'{e},
F-95302 Cergy-Pontoise, France%
}
\affiliation{
Department of Physics and Optoelectronics, Xiangtan University, Xiangtan 411105, Hunan, China
}
\date{\today}
\maketitle

\begin{figure*}[tb]
(a) \includegraphics[width=0.95\columnwidth]{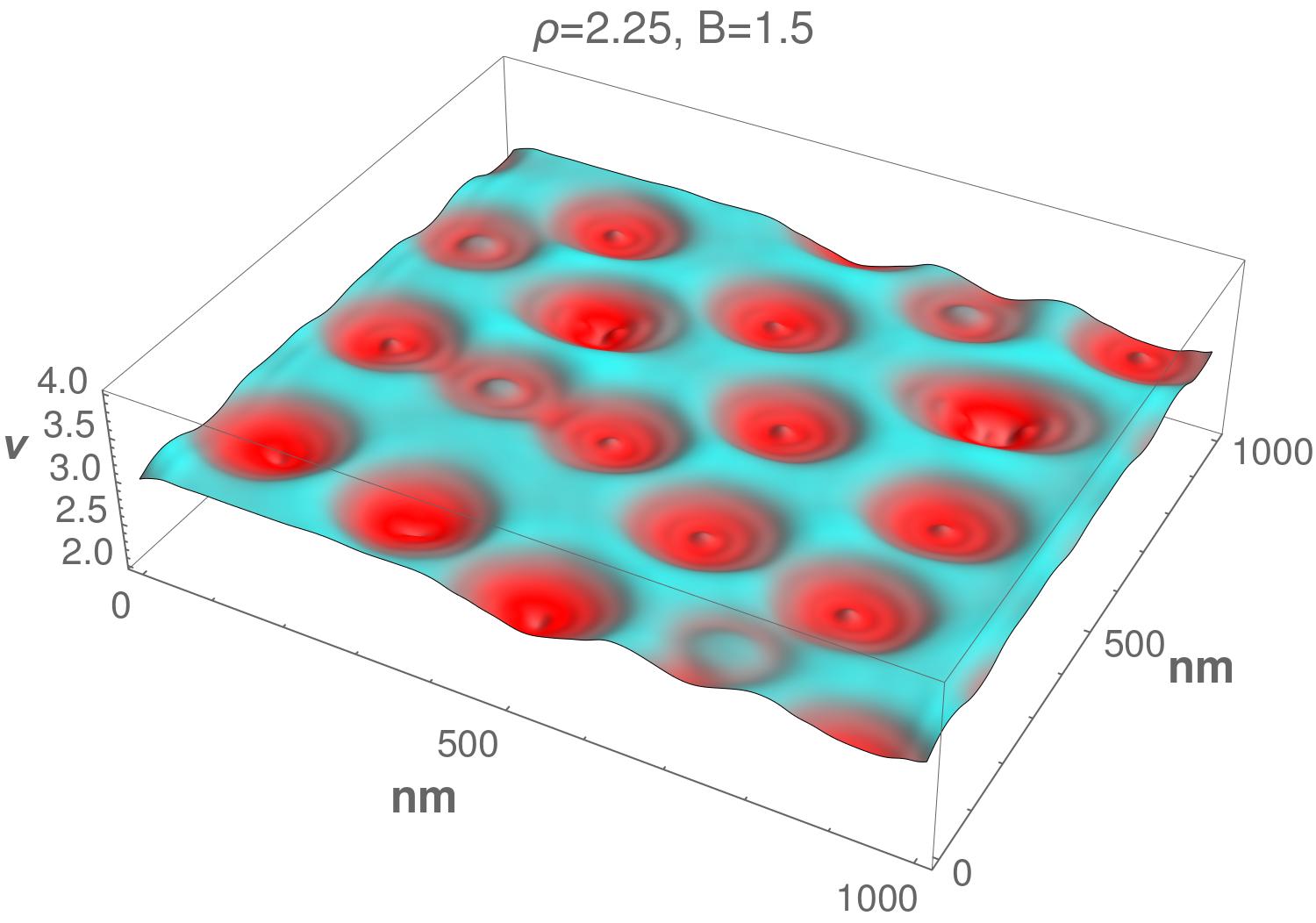}
\hspace*{-0.8\columnwidth} $\uparrow$ \hspace{0.8\columnwidth}$\downarrow$
\includegraphics[width=0.95\columnwidth]{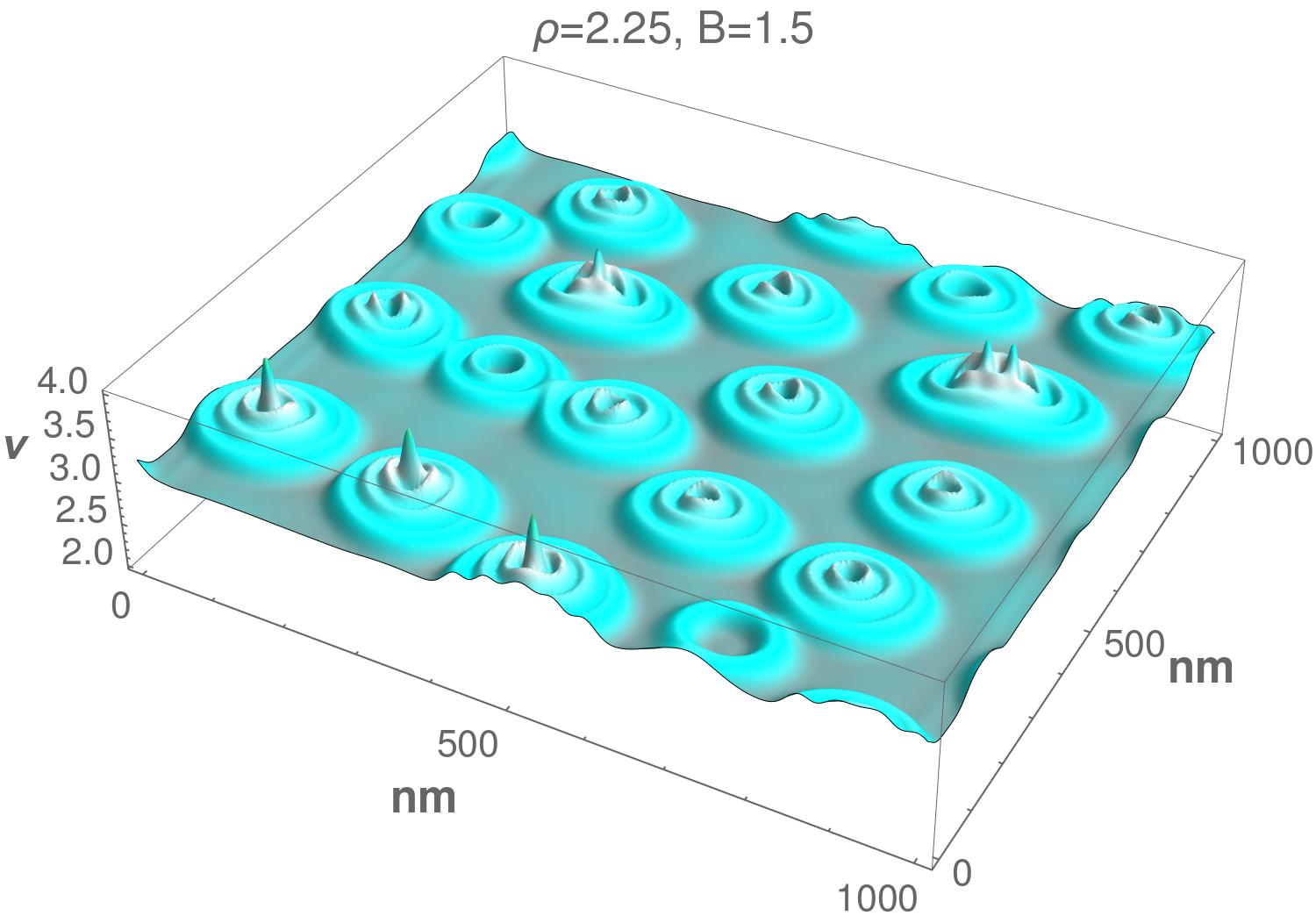}\\
(b) \includegraphics[width=0.95\columnwidth]{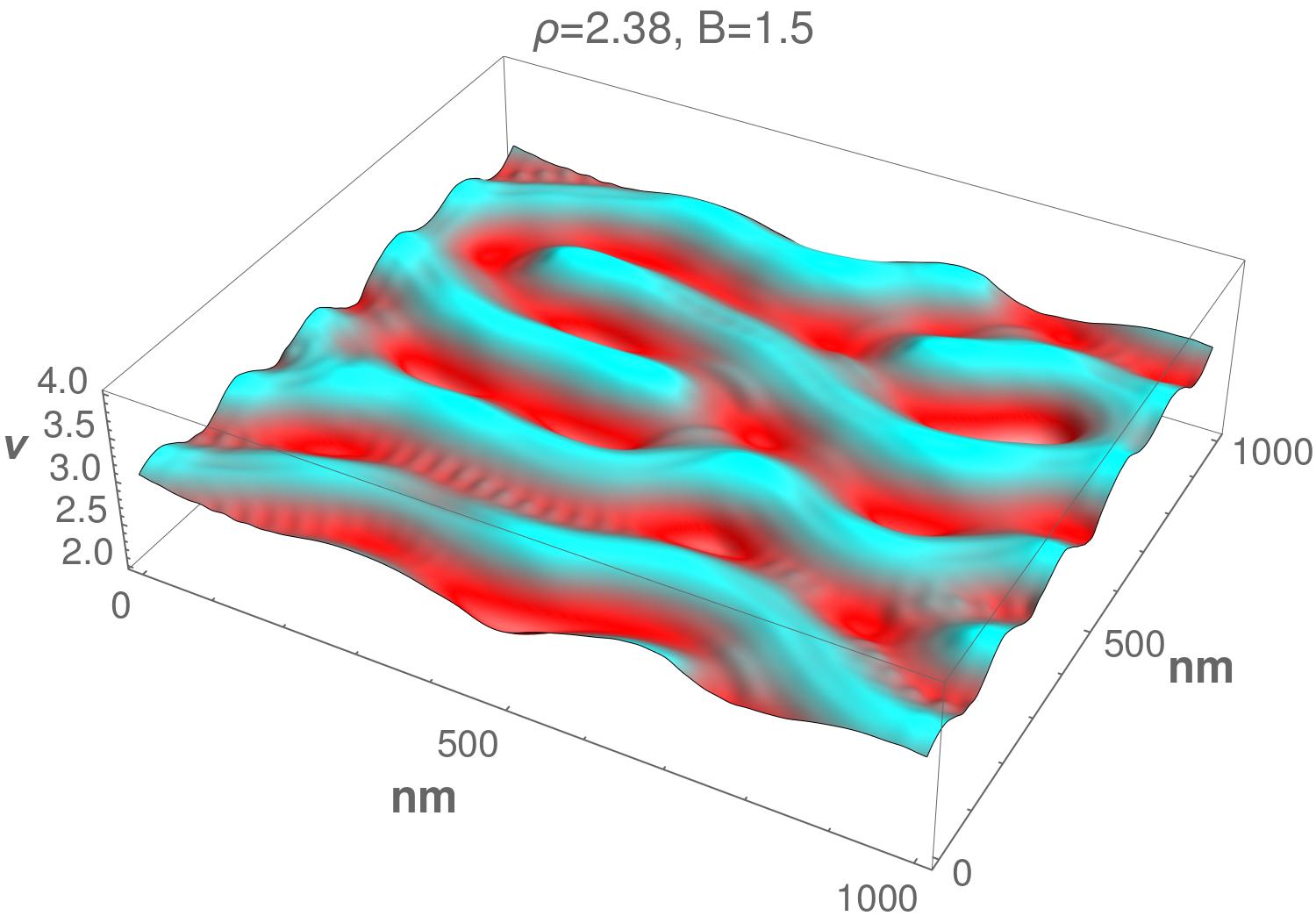}
\hspace*{-0.8\columnwidth} $\uparrow$ \hspace{0.8\columnwidth}$\downarrow$
\includegraphics[width=0.95\columnwidth]{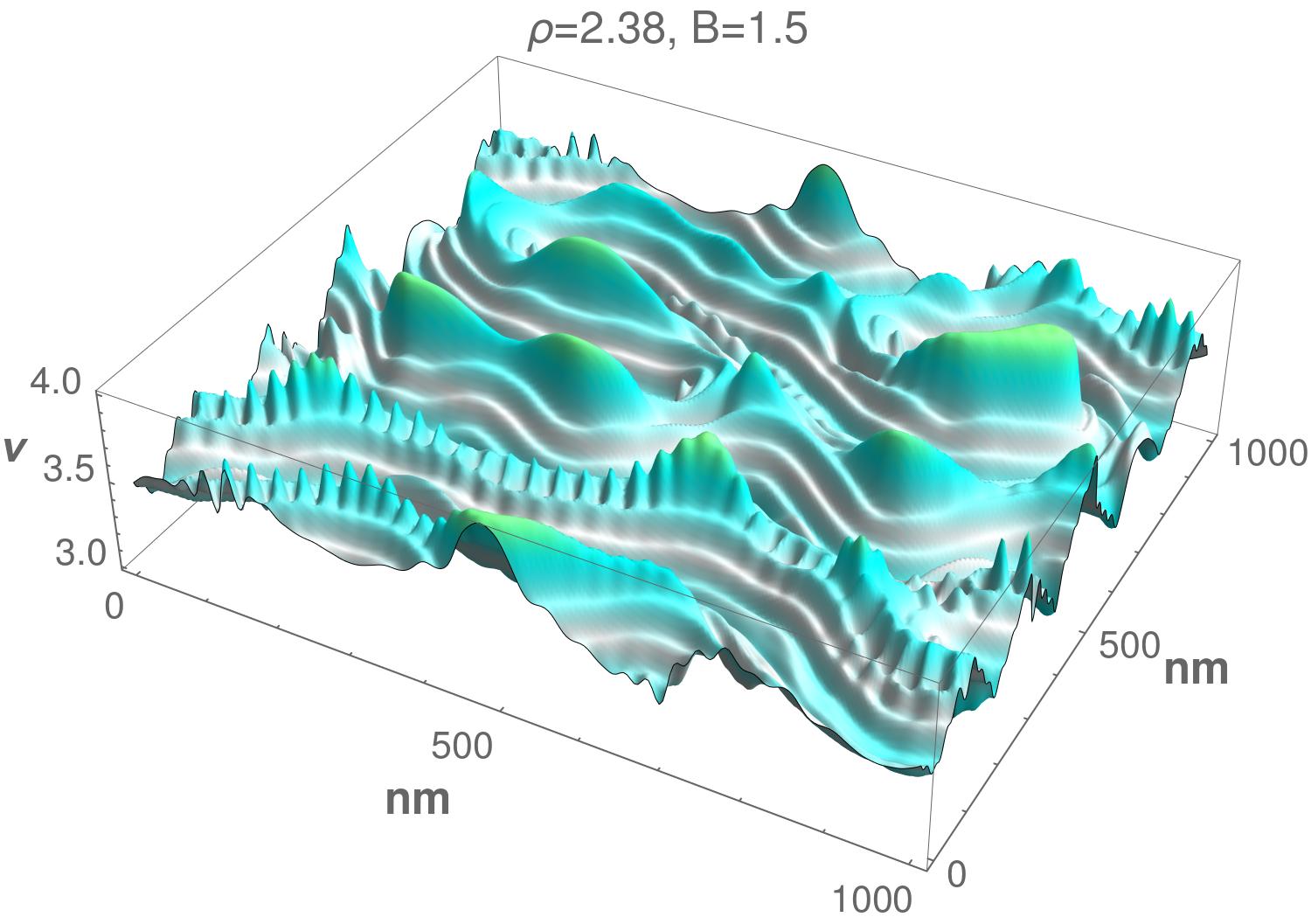}\\
(c) \includegraphics[width=0.95\columnwidth]{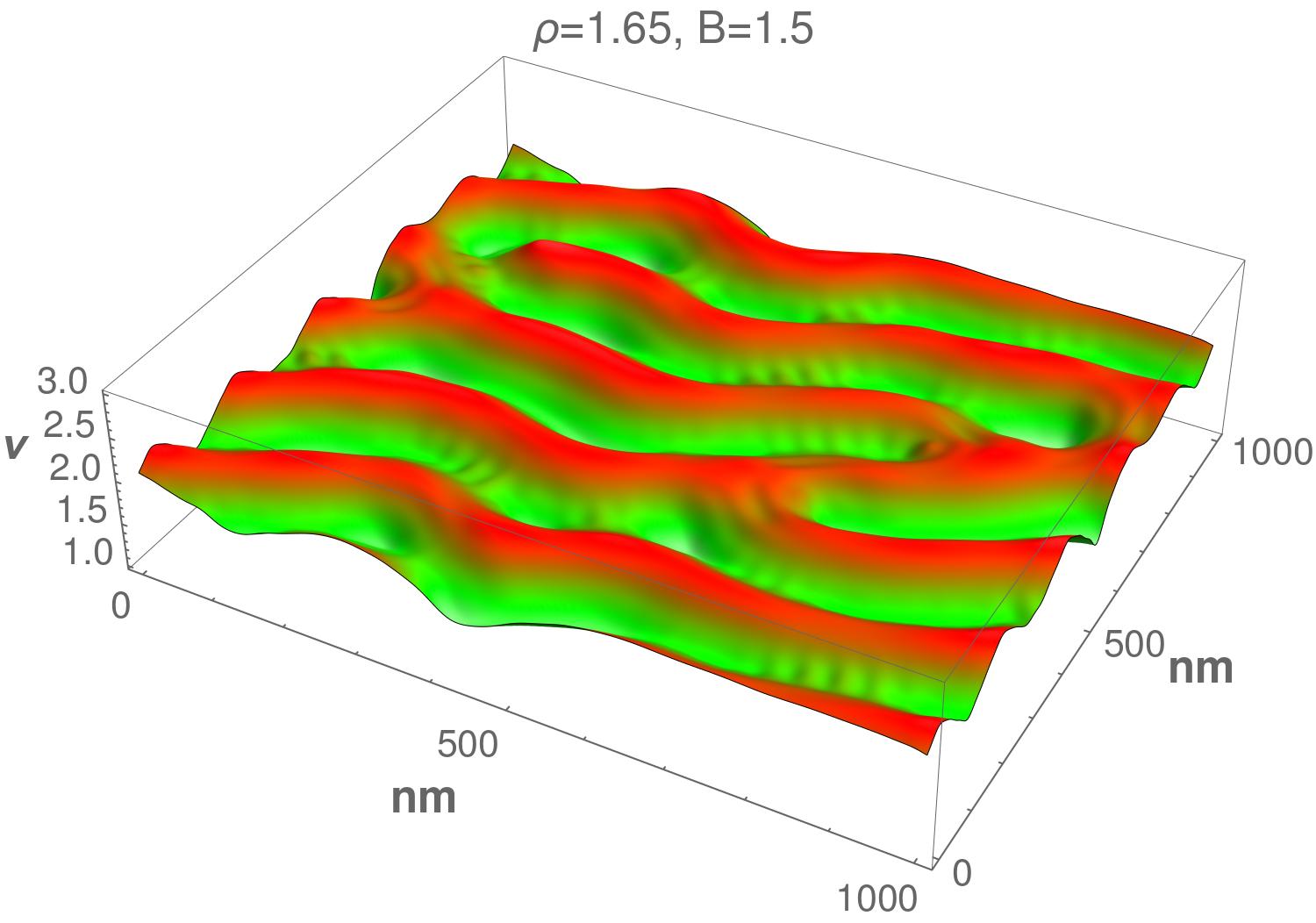}
\hspace*{-0.8\columnwidth} $\uparrow$ \hspace{0.8\columnwidth}$\downarrow$
\includegraphics[width=0.95\columnwidth]{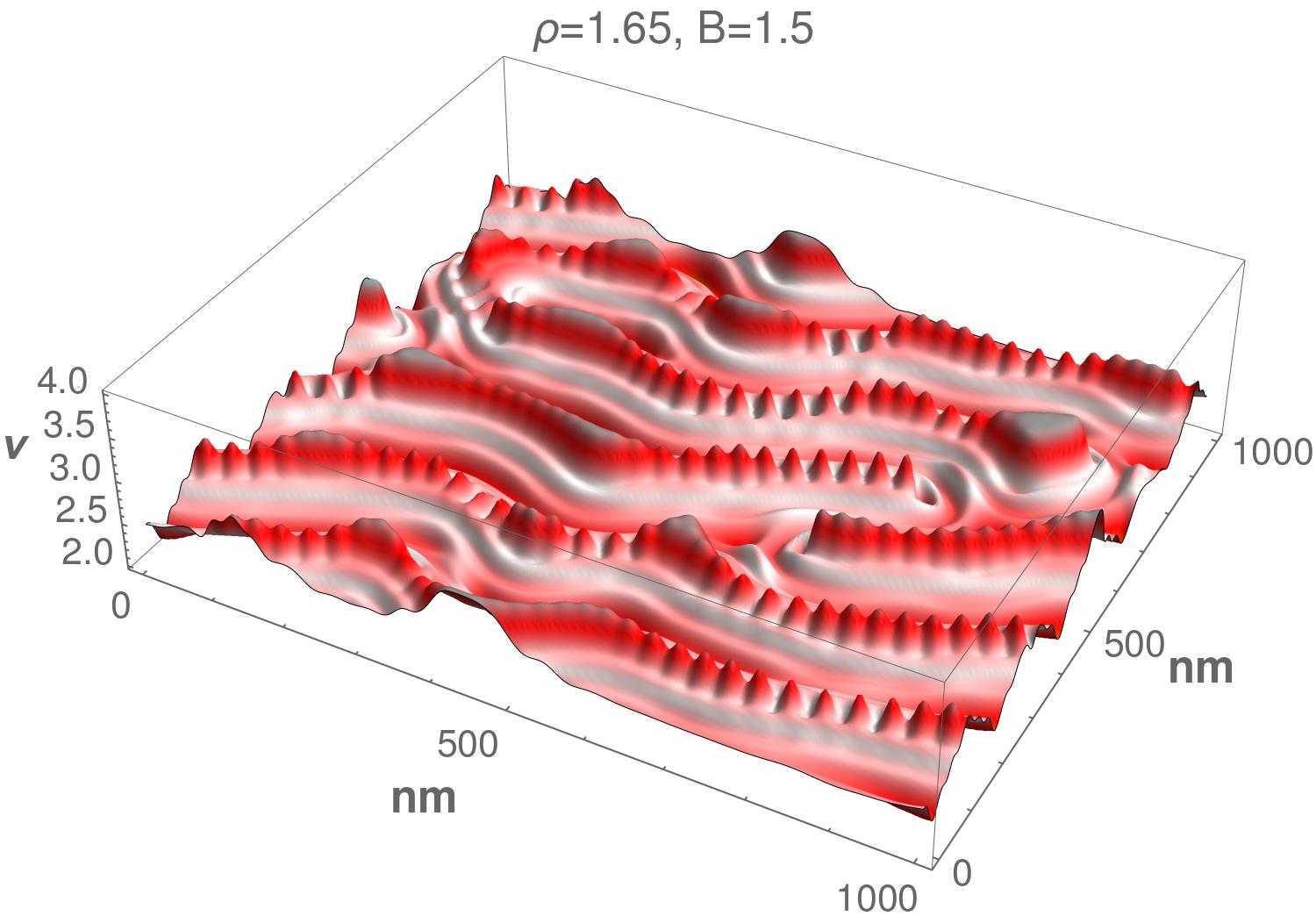}
\caption{
\label{fig-bubble_strips}
Lateral carrier density distribution mapped on the filling factor scale $\nu(\vec{r})$ for different total filling factors (a) $\nu = 6.20$, (b) $\nu = 6.54$ and (c) $\nu = 4.54$. The left and right columns contains results for $\nu_\uparrow$ and $\nu_\downarrow$, respectively. 
The color shades represent the filling factor range, where 
green indicates the second LL for $\nu_{\downarrow,\uparrow} = 1 \rightarrow 2$, red the third LL, light blue the fourth and yellow denoting LL $5$. These colours are as in Refs.\ \onlinecite{SOswaldEPL2017,SOswaldPRB2017} and the value indicated for $\rho$, $B$ denote the electron density in units of $\SI{e11}{cm^{-2}}$ and $\SI{1}{Tesla}$, respectively.
}.
\end{figure*}
\begin{figure*}[tb]
\mbox{ }  \hfill Hartree \hfill non-interacting \hfill  \mbox{ }\\
(a) $\uparrow$ \includegraphics[width=0.45\textwidth]{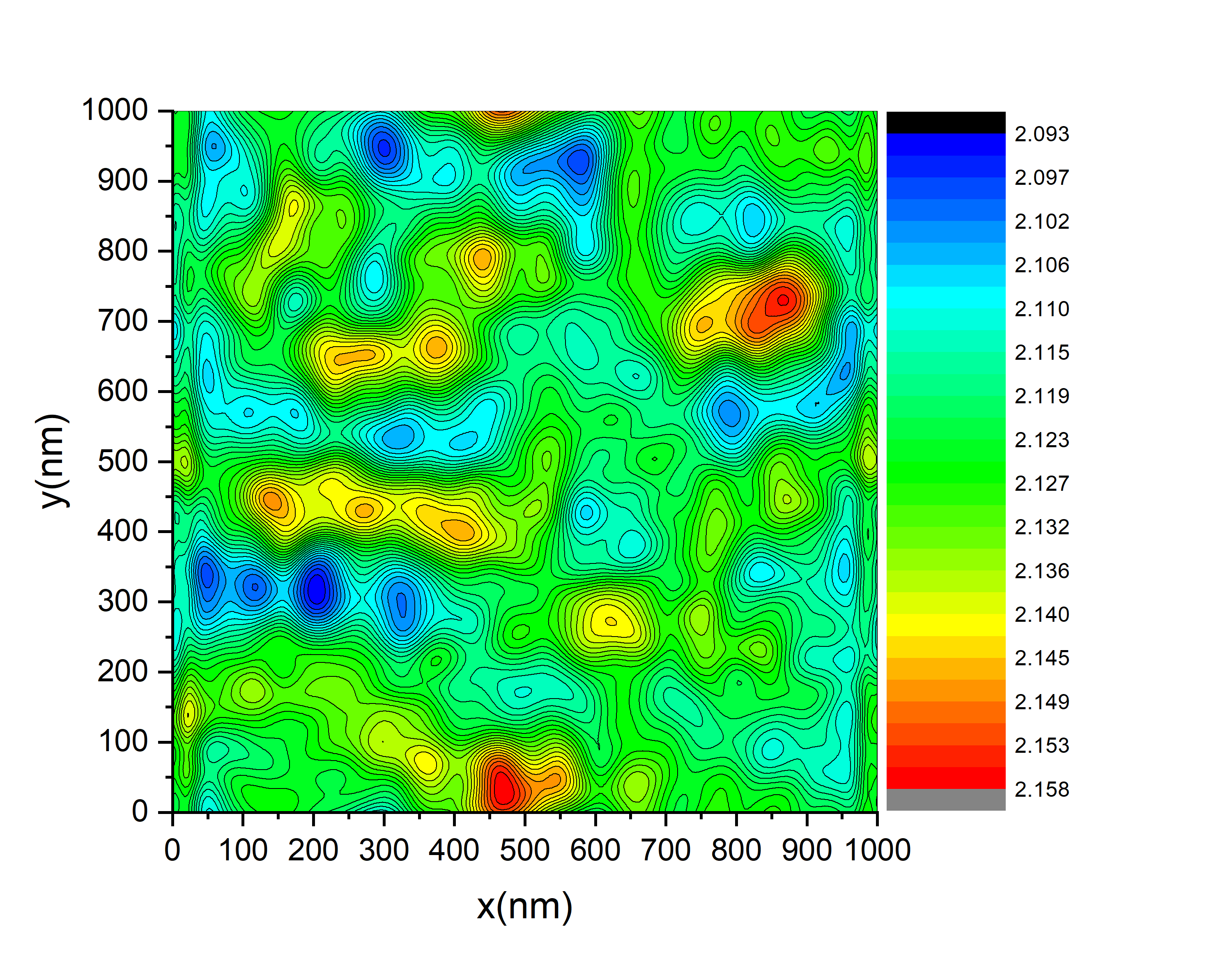}
(b) $\uparrow$\includegraphics[width=0.45\textwidth]{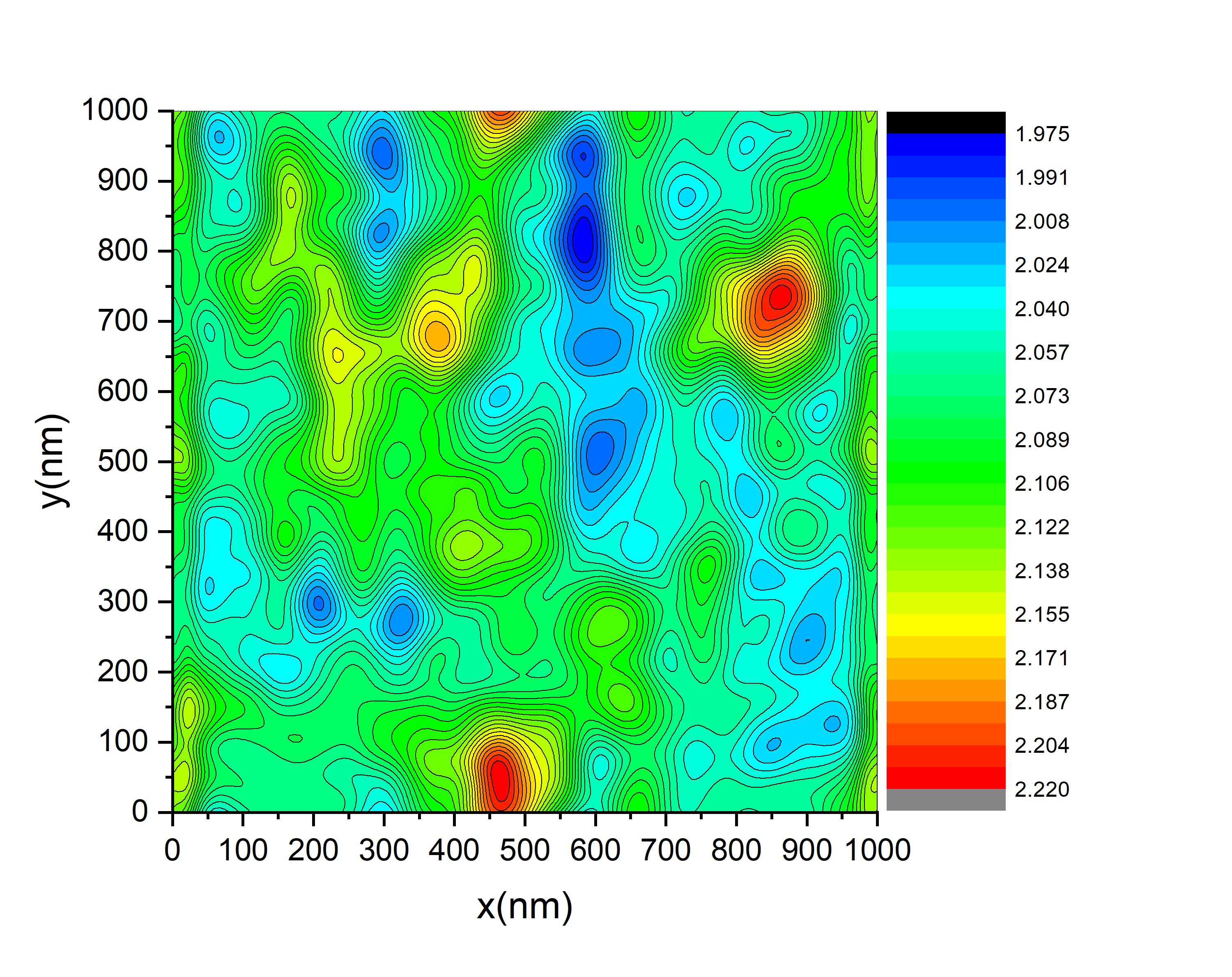}\\
(c) $\downarrow$ \includegraphics[width=0.45\textwidth]{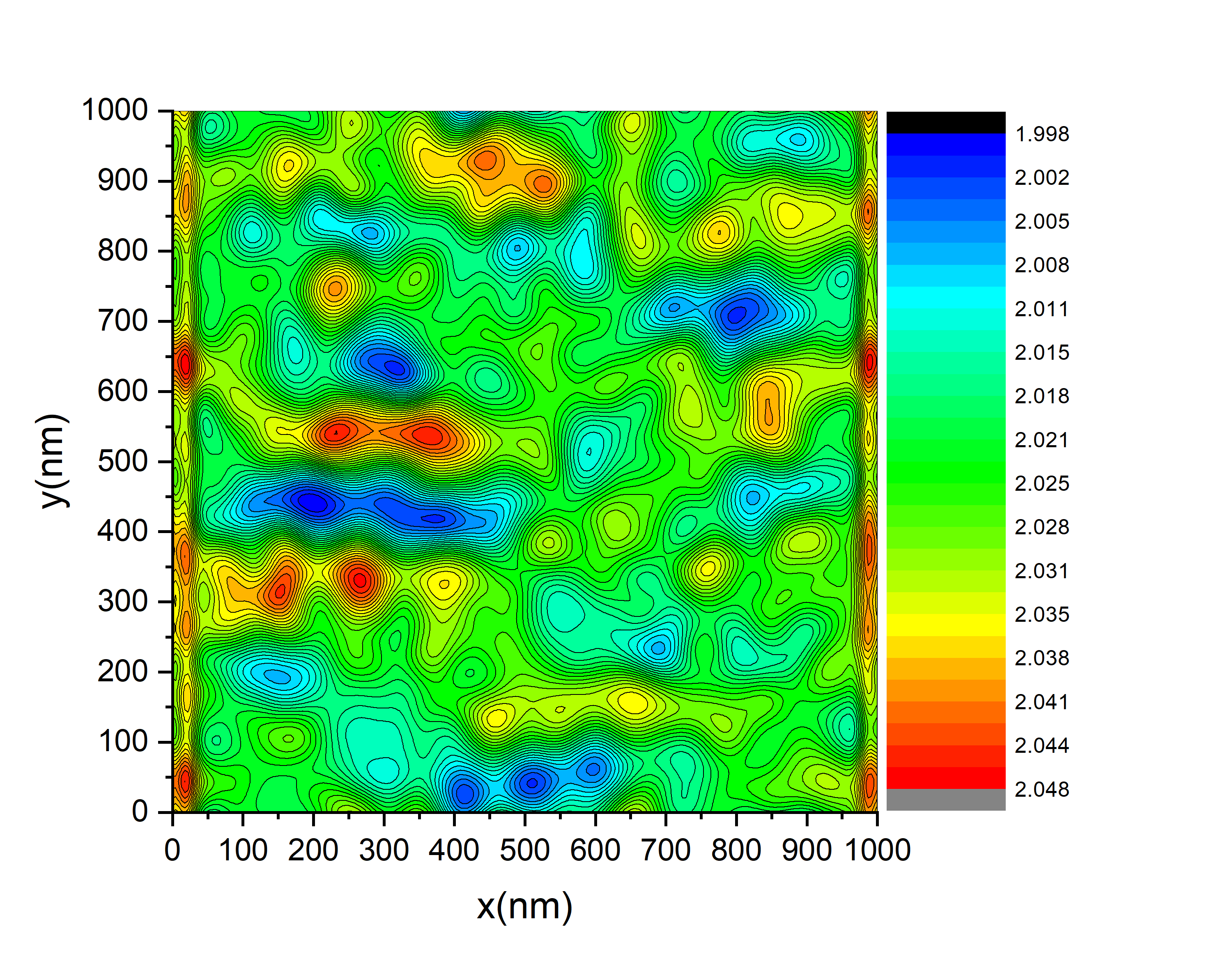}
(d) $\downarrow$\includegraphics[width=0.45\textwidth]{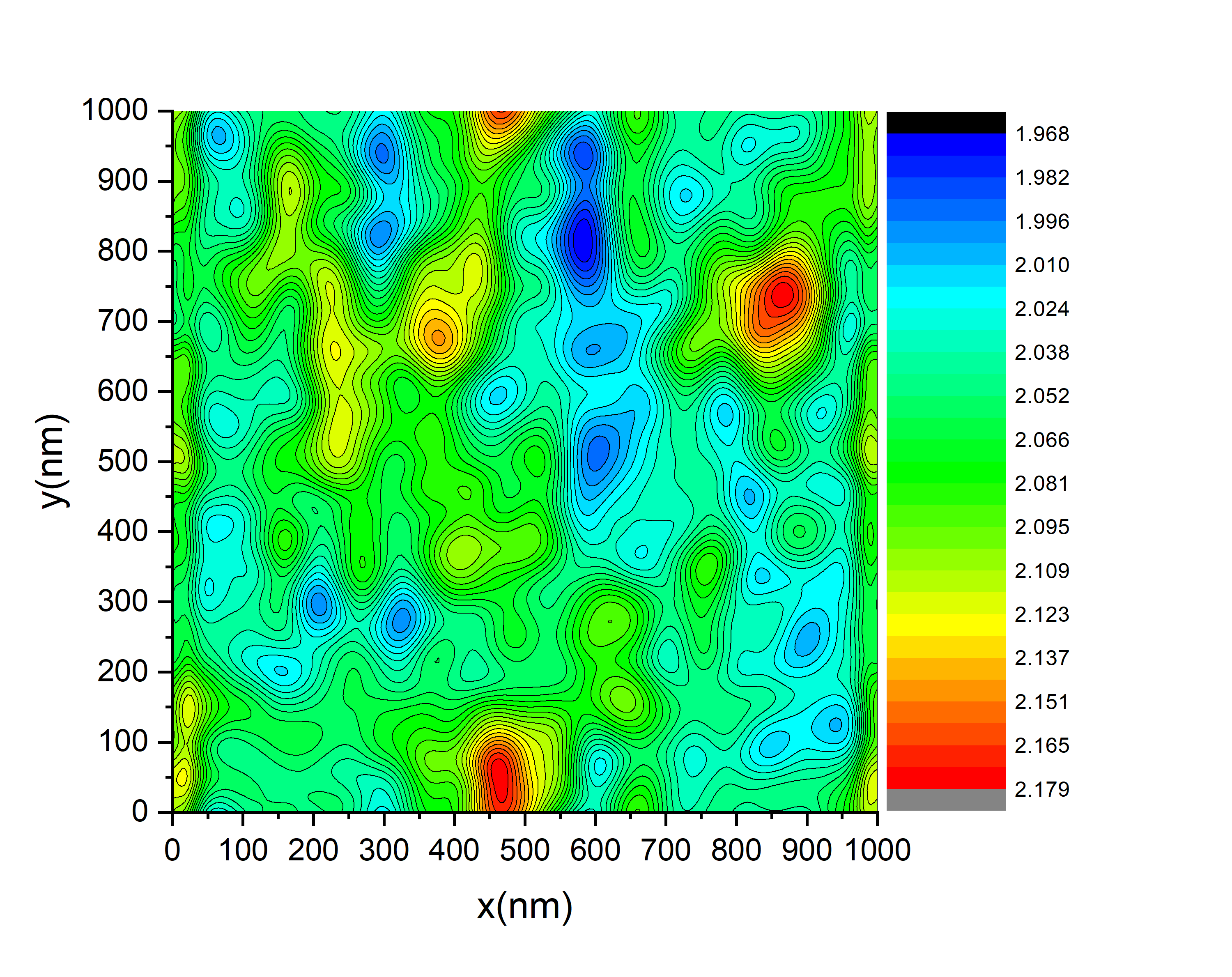}
\caption{
\label{CD_nu_450-Sx-HH-00}
Lateral charge density of the partly filled top Landau levels at total filling factor $\nu=4.5$. The top row with (a) and (b) gives the spin-down levels while to bottom row shows that spin-down levels in (c) and (d). The left column with (a) and (c) has been calculated for pure Hartree interaction while the right column with (b) and (d) shows a non-interacting situation. The color shades represent the filling factor $\nu(\vec{r})$ as given in the scales and the lines denote equal heights in $\nu(\vec{r})$.
}
\end{figure*}
\begin{figure*}[tb]
\includegraphics[width=0.95\textwidth]{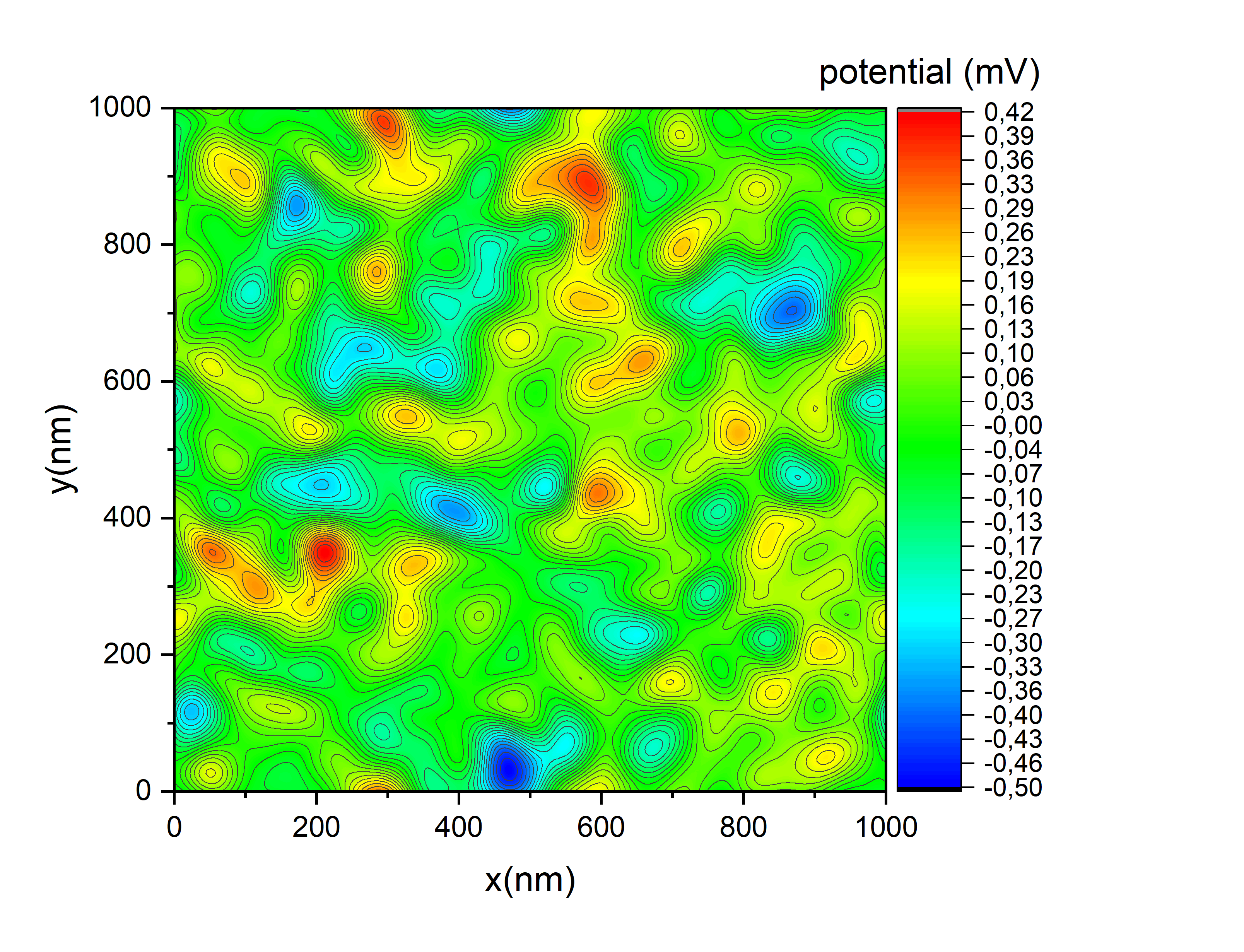}
\caption{
\label{random_pot}
Lateral random disorder potential $V(\vec{r})$ visualized in a false color plot. The lines denote equipotentials while the potential energies are indicated by the colors as in the color scale provided.
}
\end{figure*}

In Fig.\ \ref{fig-bubble_strips} we show the variation in $\nu(\vec{r})$ for three different densities at fixed magnetic field $B$. These results are identical to Fig.\ 1 of the main text for $\nu_{\downarrow}(\vec{r})$ but also show $\nu_{\uparrow}(\vec{r})$ in the left column. The choice of colors is as in Refs.\ \cite{SOswaldEPL2017,SOswaldPRB2017}.
In Fig.\ \ref{CD_nu_450-Sx-HH-00}(a) and (c) one can see that for pure Hartree interaction there is no stripe formation. The density modulation in $\nu(\vec{r})$ is much less than in Figs.\ \ref{fig-nur} and \ref{fig-bubble_strips} and roughly follows the random potential shown in Fig.\ \ref{random_pot}. Furthermore, the charge density modulation in the spin-up and spin-down levels "repel" each other due to the Hartree interaction.
In Fig.\ \ref{CD_nu_450-Sx-HH-00}(b) and (c) we show the situation without interaction. Clearly, there is also no stripe formation. The $\nu(\vec{r})$ modulation rather closely follows the random potential of Fig.\ \ref{random_pot}.  The charge densities in the spin-up and spin-down levels do not influence each other and follow nearly identically the disorder potential because of any missing interaction.

\begin{figure*}[tb]
\includegraphics[width=0.95\textwidth]{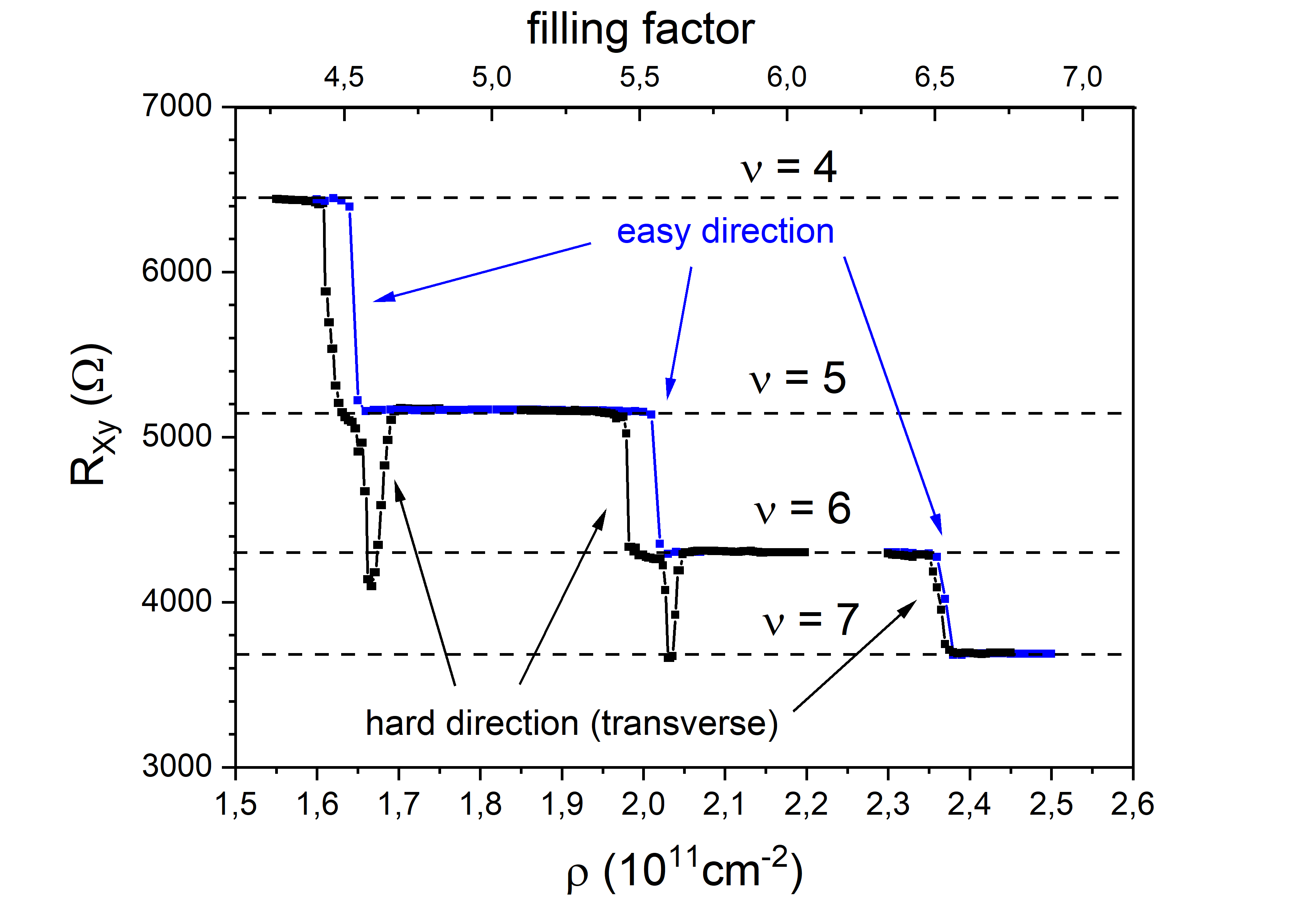}
\caption{
\label{transport_Rxy}
Hall resistance $R_{xy}$ for the longitudinal resistance $R_{xx}$ shown in Fig.\ 2 with the same set of parameters.
}
\end{figure*}

Fig.\ \ref{transport_Rxy} complements Fig.\ \ref{transport} by showing in addition the Hall resistance $R_{xy}$.

\begin{figure*}[tb]
\includegraphics[width=0.95\textwidth]{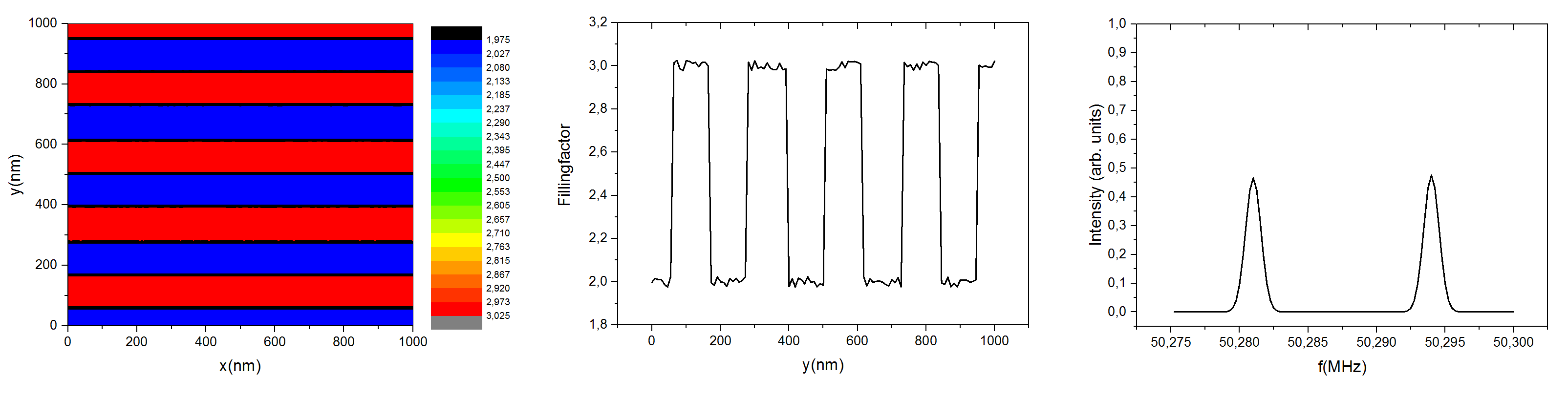}
\includegraphics[width=0.95\textwidth]{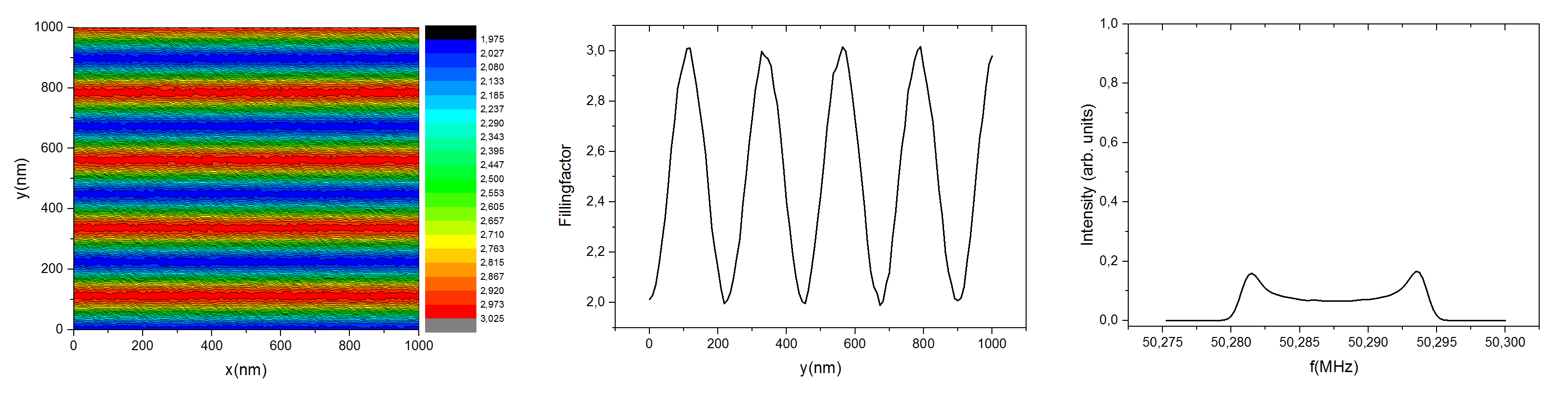}
\includegraphics[width=0.95\textwidth]{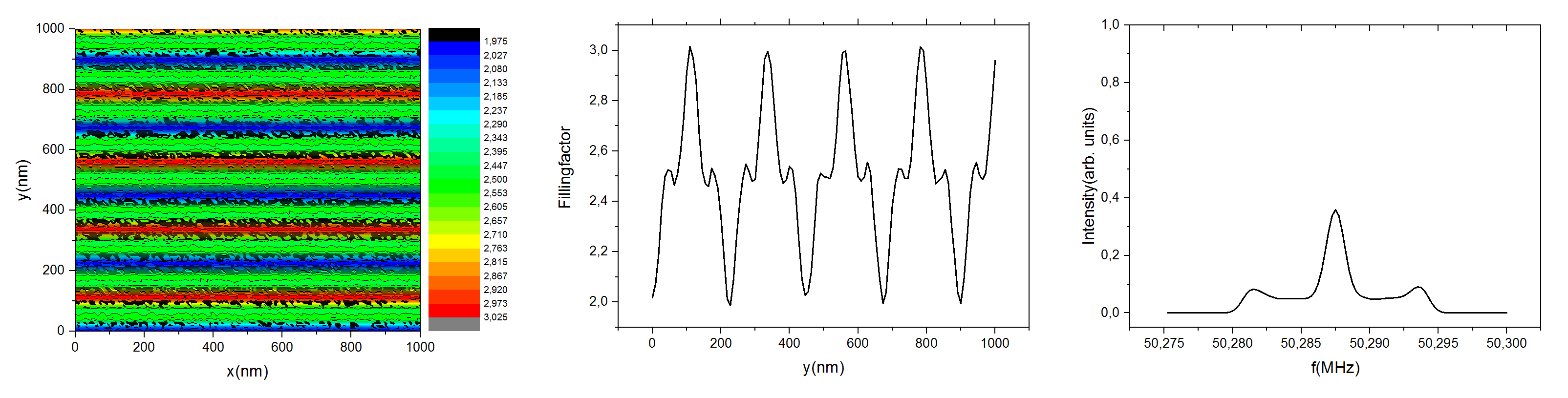}
\includegraphics[width=0.95\textwidth]{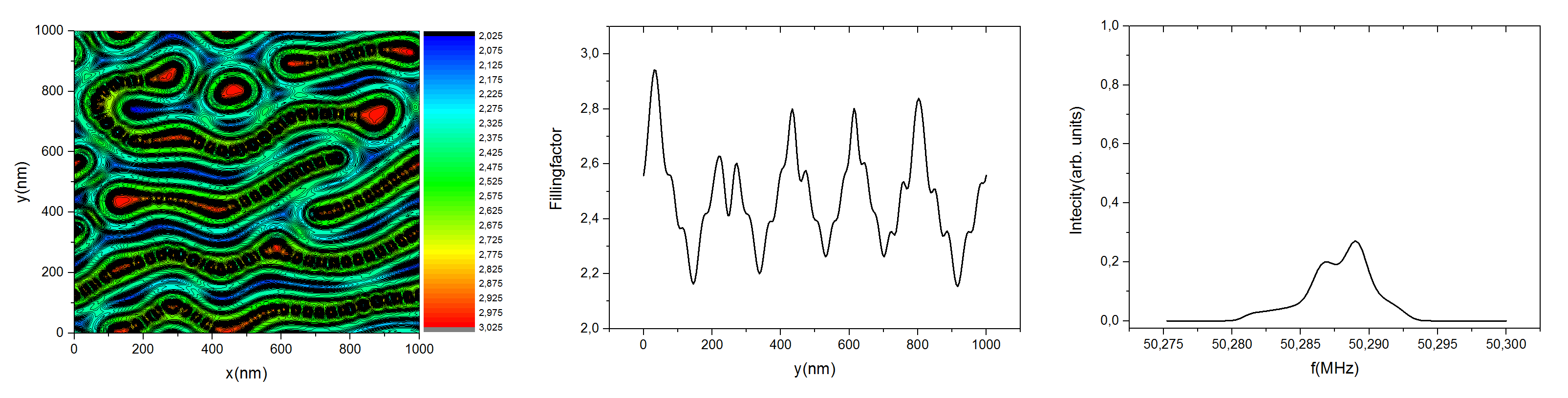}
\caption{
\label{fig-NMR-test}
Calculations of the NMR intensity $I_{\nu}(f)$ for 4 different stripe-like variations of $\nu(\vec{r})$ at $\nu=4.5$. Rows 1-4 corresponds to (a) a simply square modulation, (b) a sinusoidal modulation, (c) a sinusoidal variations with left and right shoulders, (d) a modulation as computed from HF. The first column shows the spatial variations in $\nu(\vec{r})$ as given by the color scales. The second column represent a typical cross-section for each situation and the third column shows the estimated NMR intensity $I_{4.5}(f)$.
}.
\end{figure*}

In order to show that the presence of the remnants of the LL wave functions around each stripe is significant, we perform the calculations of $I_{\nu}(f)$ for three test patterns for $\nu(\vec{r})$. The results are given in Fig.\ \ref{fig-NMR-test}. We find that only the variation given by $\nu(\vec{r})$ as calculated in HF can reproduce essential global features of the experimental NMR results presented in Ref.\ \cite{SFriess2014}. 

\begin{figure*}[tb]
\mbox{ }  \hfill $\nu=2$--$3$ \hfill $\nu=4$--$5$ \hfill  \mbox{ }\\
(a)\includegraphics[width=0.47\textwidth]{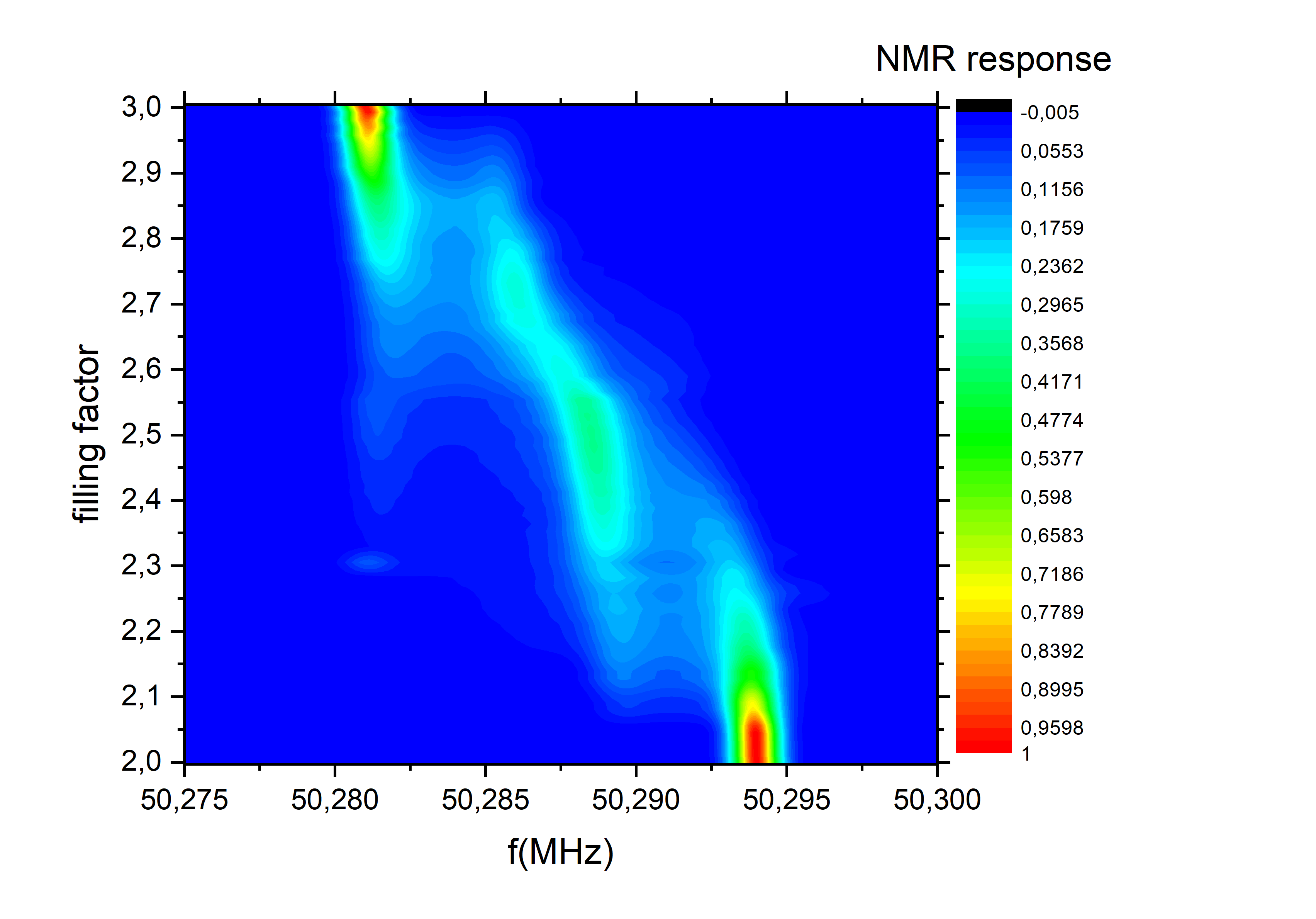}
(b)\includegraphics[width=0.47\textwidth]{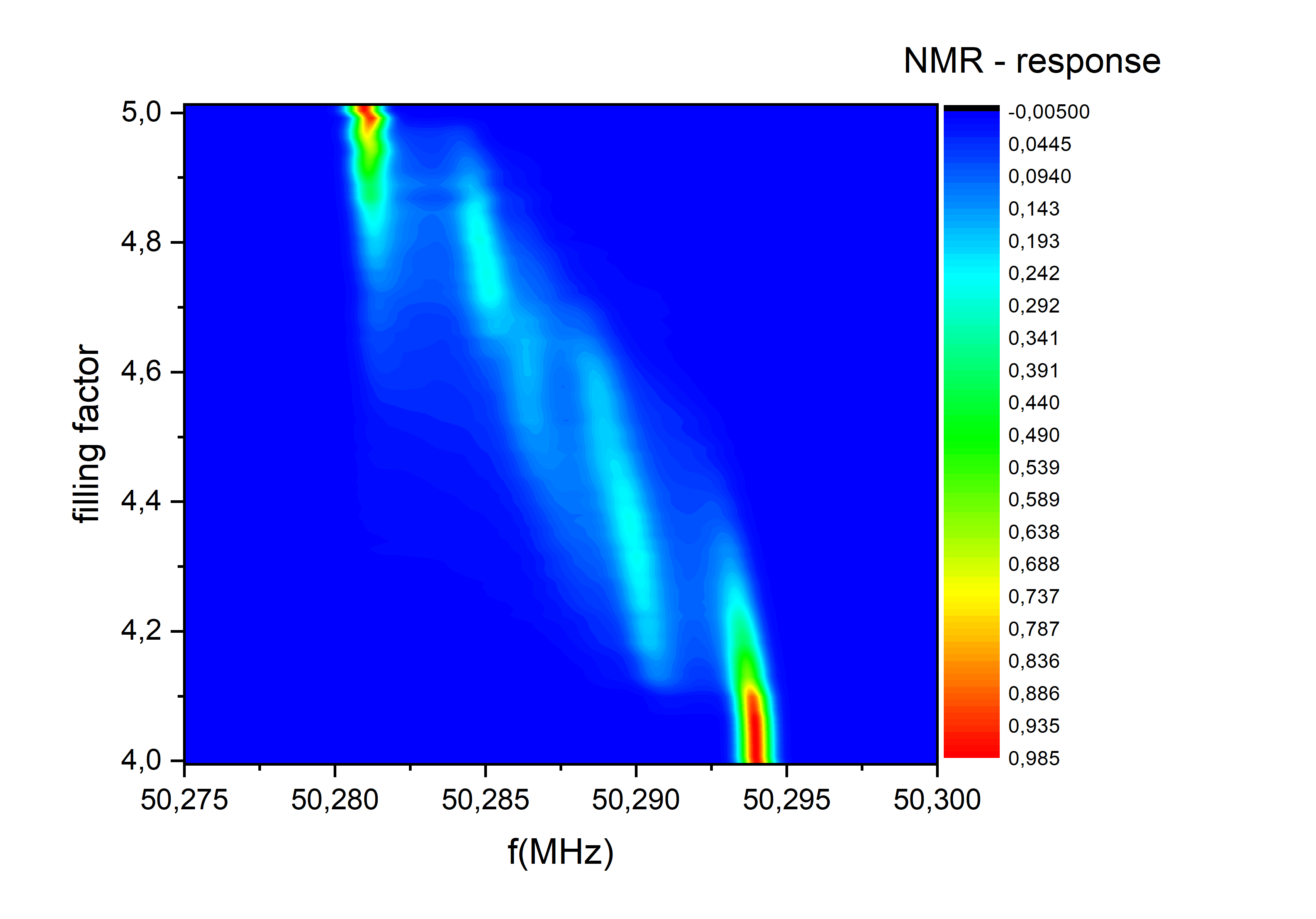}\\
\mbox{ }  \hfill $\nu=2.5$ \hfill $\nu=4.5$ \hfill \mbox{ }\\
(c)\includegraphics[width=0.47\textwidth]{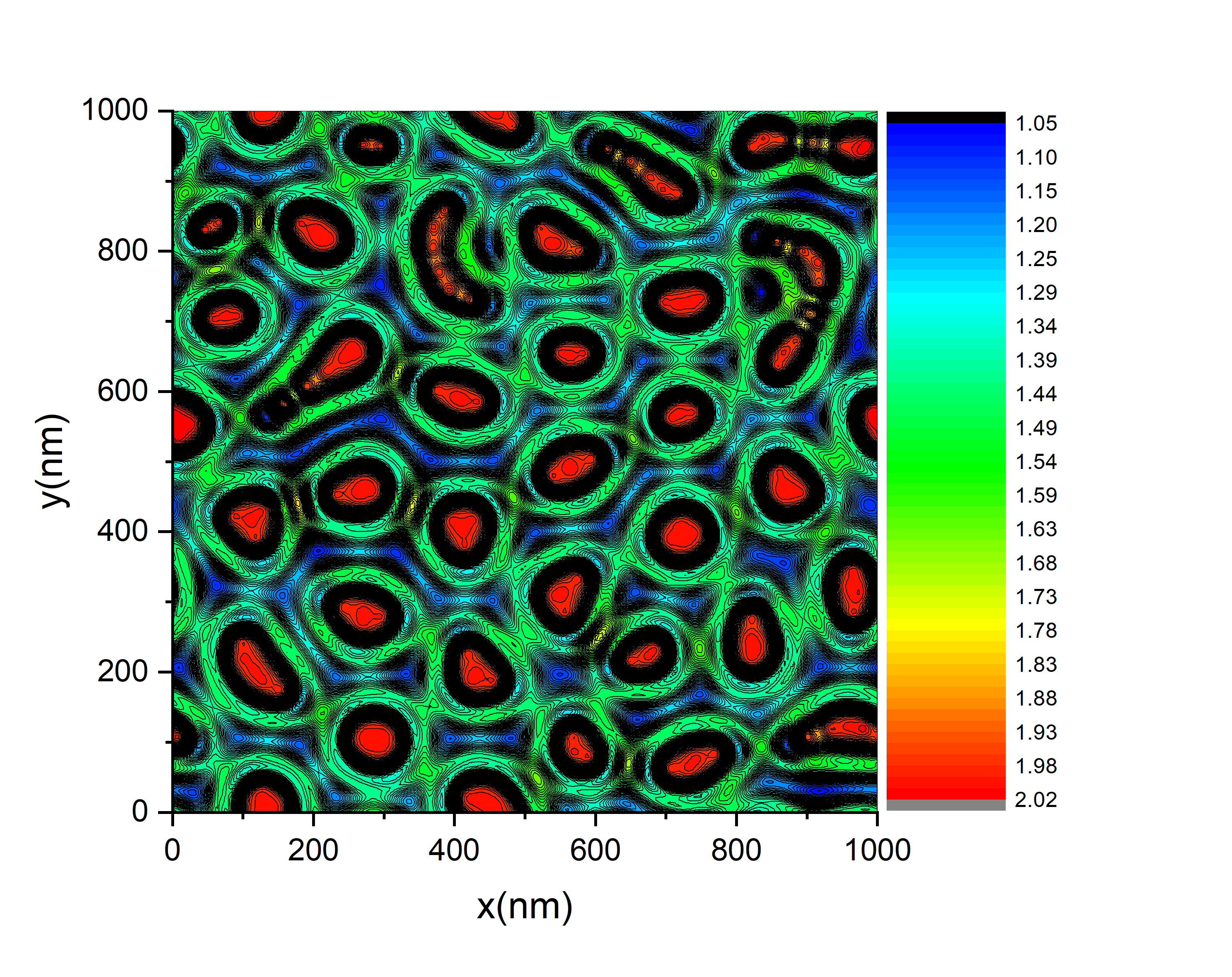}
(d)\includegraphics[width=0.47\textwidth]{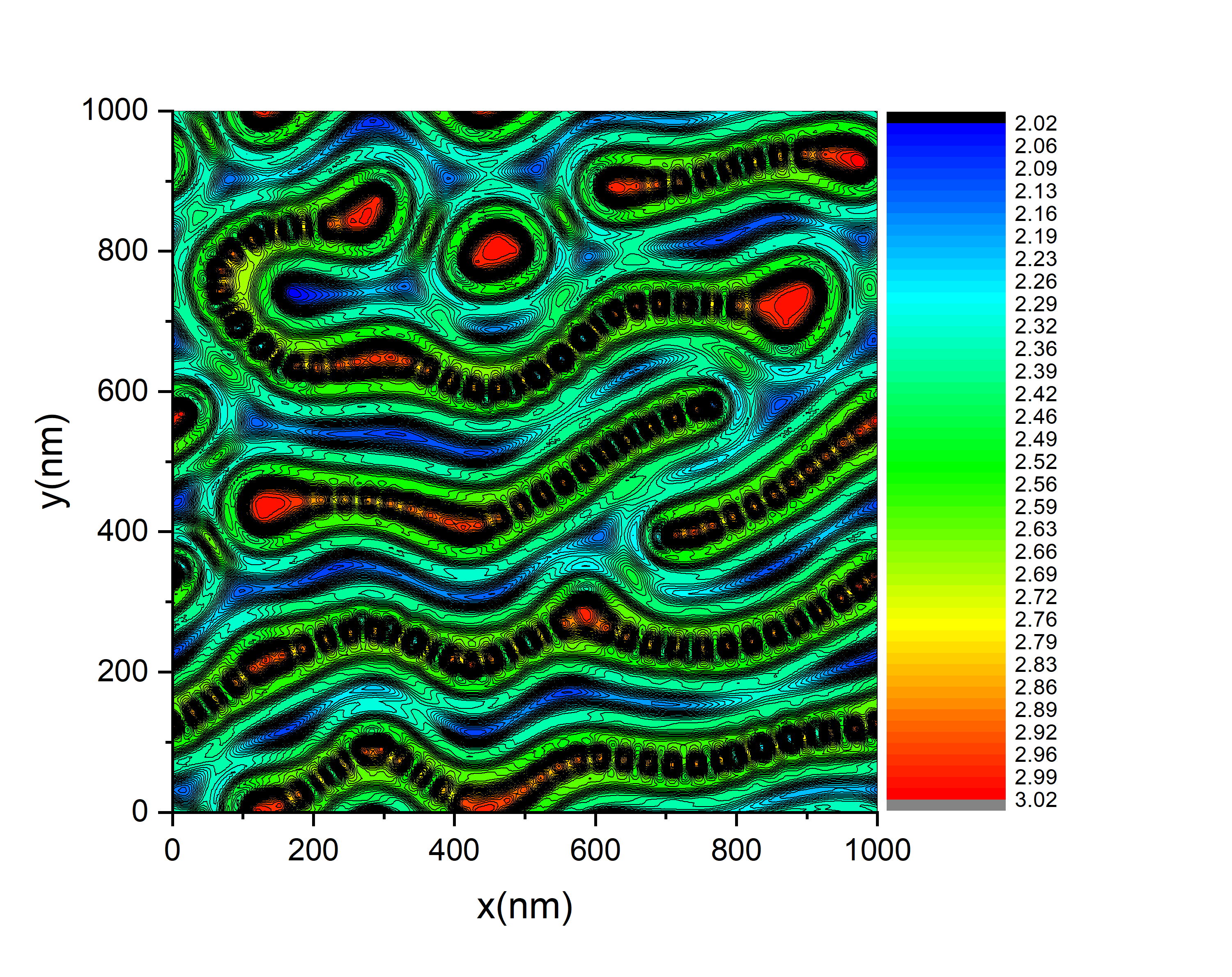}
\caption{
\label{fig-NMR-stripes-bubbles}
NMR intensities $I_{\nu}(f)$ for (a) $\nu=2-3$ and (b) $\nu=4-5$ and local $\nu(\vec{r})$ at (c) $\nu=2.5$ and (d) $\nu=4.5$ in left and right columns, respectively. (b+d) The right column reproduces results already shown in Figs.\ \ref{fig-nur} and \ref{fig-NMR}. The color shades represent  $I_{\nu}(f)$ and $\nu(\vec{r})$ as given by the scales. Lines in (c+d) connect equal height in $\nu(\vec{r})$.
}
\end{figure*}

\begin{figure*}[tb]
\includegraphics[width=0.95\textwidth]{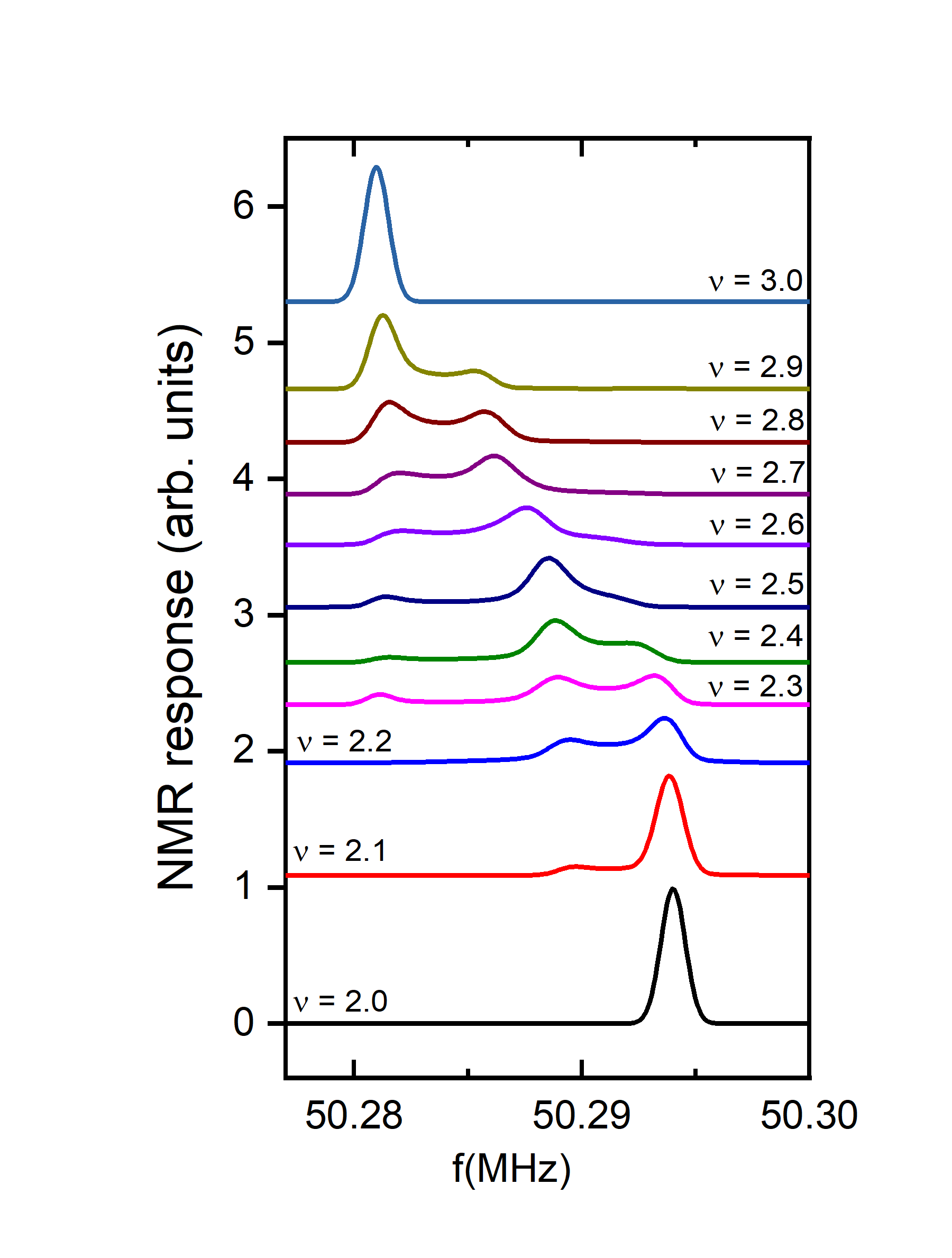}
\caption{
\label{fig:NMRspec-2-3}
NMR intensities $I_{\nu}(f)$ as shown in Fig.\ \ref{fig-NMR}(a) but for the filling factor range $\nu=2 - 3$.
}
\end{figure*}

In Fig.\ \ref{fig-NMR-stripes-bubbles} we show the behavior of $I_{\nu}(f)$  for stripes and bubble-like charge density waves. As in Fig.\ \ref{fig-NMR-test}, the HF results for $\nu(\vec{r})$ lead to a reasonable qualitative agreement with the non-interacting model used in Ref.\ \cite{SFriess2014}. Nevertheless, the details around, e.g., $\nu=2.5$ are rather different, highlighting the importance of interactions.
We note that normally stripes appear only starting with filling factor $\nu=4.5$. This is known also experimentally, but for experimental reasons the authors of Ref.\ \onlinecite{SFriess2014} could not go to that filling factor. Instead, they used filling factor $\nu=2.5$ and forced, by using an in-plane component of the magnetic field, the electron system to form a stripe pattern. Clearly, there is no need for our simulations to also model this experimental "trick". 
In order to compare the effect of stripe patterns on the NMR Knight shift, we therefore use  the stripe pattern in the "correct" range $\nu=4$--$5$. The result in Fig.\ \ref{fig-NMR-stripes-bubbles} (b) has striking similarities but seems indeed a bit richer in features than the experimental curve for $\nu=2$--$3$. 
However, since the Knight shift spectrum looses its local information due to the spatial integration, we can also evaluate the range $\nu = 2$--$3$ as shown in Fig.\ \ref{fig-NMR-stripes-bubbles}. Indeed the agreement with the experiments of Ref.\ \onlinecite{SFriess2014} becomes even better in this filling factor range. We can still see in total 3 peaks, two of them clearly separated and a third one as a shoulder on the high frequency flank, just as shown for the experiments in Fig. 2b of Ref.\ \onlinecite{SFriess2014}. This makes the agreement almost perfect for the experiments as shown in Fig.\ \ref{fig:NMRspec-2-3} albeit not with the non-interacting modelling of Fig.\ 2c \cite{SFriess2014}. Filling factor $6$--$7$ requires much more computing time and, although a most interesting question, there are currently no experiments available for comparison.



{}

\begin{thebibliography}{44}
\bibitem{scale}
{The scale  $\SI{1/1000}{nm^{-1}}$ of the reciprocal period is chosen in a way that it matches the number of periods appearing on the length of a sample square of $1000 \times \SI{1000}{nm^2}$.}
\end{thebibliography}

\end{document}